\documentclass[lettersize,journal]{IEEEtran}
\usepackage{amsmath,amsfonts, bm}
\usepackage{array}
\usepackage[caption=false,font=normalsize,labelfont=sf,textfont=sf]{subfig}
\usepackage{textcomp}
\usepackage{amssymb}
\usepackage{amsthm}
\usepackage{stfloats}
\usepackage{url}
\usepackage{verbatim}
\usepackage{graphicx}
\usepackage{cite}
\usepackage{nicefrac}
\usepackage{subfig, caption}
\usepackage{xcolor}
\usepackage{multicol, multirow}
\usepackage{algorithm2e} 
\usepackage{booktabs}
\usepackage{cuted}

\newcommand{\real}{\mathbb{R}}

\newcommand{\diag}{\mathrm{diag}}
\newcommand{\A}{\mathbf{A}}
\newcommand{\E}{\mathbf{E}}
\newcommand{\e}{\mathbf{e}}
\newcommand{\x}{\mathbf{x}}
\newcommand{\X}{\mathbf{X}}
\newcommand{\ba}{\mathbf{a}}
\newcommand{\bS}{\mathbf{S}}
\newcommand{\s}{\mathbf{s}}

\newcommand{\z}{\mathbf{z}}

\newtheorem{theorem}{Theorem}

\newtheorem{definition}{Definition}

\SetAlCapFnt{\small}
\SetAlCapNameFnt{\small\bfseries}
\DontPrintSemicolon
\LinesNumbered 

\begin{document}

\title{A Tractable Two-Step Linear Mixing Model Solved with Second-Order Optimization for Spectral Unmixing under Variability}

\author{Xander Haijen, Bikram Koirala, Xuanwen Tao, and Paul Scheunders
\thanks{Xander Haijen, Bikram Koirala, Xuanwen Tao, and Paul Scheunders are with the Imec-Visionlab research group, University of Antwerp.}
\thanks{The research presented in this paper is funded by the Research Foundation-Flanders - project G031921N. Xander Haijen is a PhD fellow of the Research Foundation Flanders, Belgium (FWO: 1103226N). Bikram Koirala is a postdoctoral fellow of the Research Foundation Flanders, Belgium (FWO: 1250824N-7028).}%
}

\markboth{}%
{Haijen \MakeLowercase{\textit{et al.}}: 2LMM: a Tractable Two-Step Linear Mixing Model Solved with Second-Order Optimization for Spectral Unmixing under Variability}


\maketitle

\begin{abstract}
Spectral unmixing is an important task in hyperspectral image processing. It can be thought of as a regression problem, where the observed variable (i.e., an image pixel) is to be found as a function of the response variables (i.e., the pure materials in a scene, called endmembers). The Linear Mixing Model (LMM) has received a great deal of attention, due to its simplicity and ease of use in, e.g., optimization problems. Its biggest flaw is that it assumes that any pure material can be characterized by one unique spectrum throughout the entire scene. In many cases, this is incorrect: the endmembers face a significant amount of spectral variability caused by, e.g., illumination conditions, atmospheric effects, or intrinsic variability. Researchers have suggested several generalizations of the LMM to mitigate this effect. However, most models lead to ill-posed and highly non-convex optimization problems, which are hard to solve and have hyperparameters that are difficult to tune. They also use first-order optimization methods, which are not suitable for the complex problems they address.
In this paper, we propose a Two-Step Linear Mixing Model (2LMM) that bridges the gap between model complexity and computational tractability. The model achieves this by introducing two distinct scaling steps: an endmember scaling step across the image, and another for pixel-wise scaling. We show that this model leads to only a mildly non-convex optimization problem, which we solve with an optimization algorithm that incorporates second-order information. To the authors' knowledge, this work represents the first application of second-order optimization techniques to solve a spectral unmixing problem that models endmember variability. Our method is highly robust, as it requires virtually no hyperparameter tuning and can therefore be used easily and quickly in a wide range of unmixing tasks. We show through extensive experiments on both simulated and real data that the new model is competitive and in some cases superior to the state of the art in unmixing. The model also performs very well in challenging scenarios, such as blind unmixing.
\end{abstract}


\section{Introduction}
\IEEEPARstart{H}{yperspectral} imaging  (HSI) has been widely used as an alternative to high-spatial-resolution RGB images in remote sensing for the detection of, e.g., terrestrial features and ground cover classification \cite{shaw_spectral_2003}. As HSI typically lacks sufficient spatial resolution, an important task is \textbf{spectral unmixing}, i.e., the subpixel estimation of the coverages of pure materials. A major challenge during unmixing is the fact that the signatures of the pure materials vary throughout the image, due to various causes. As a result, researchers have focused much attention on developing unmixing methods that can mitigate this effect.



\subsection{Unmixing under spectral variability} \label{sec: unmixing under variability}

The linear mixing model (LMM) serves as the basis for many unmixing models and techniques. The basic assumption of the LMM is that the surface within a given image is covered by a low number of distinct pure materials that have relatively constant spectral signatures. These distinct materials are called the \textit{endmembers} (EMs) \cite{keshava_spectral_2002}. The LMM assumes that the reflected spectrum in each pixel can be described as a linear combination of the pure materials. The relative area covered by any given EM is known as its \textit{fractional abundance}. 
Performing unmixing using the LMM is usually done by minimizing the Euclidean distance between the measured spectrum and a reconstructed spectrum. This leads to a convex minimization problem, which is separable per pixel. As a consequence, the problem can be solved efficiently and in a parallelized way.

However, the LMM is not suited for performing unmixing when spectral variability is present. It has a significant model mismatch, which leads to poor results. As a result, researchers have developed a wide variety of methods to mitigate the effects of spectral variability \cite{borsoi_spectral_2021}. In these methods, EM signatures either originate from a spectral library or can be obtained from the image using an Endmember Extraction Algorithm (EEA). 
The most important unmixing methods are summarized below.

\subsubsection{Methods that use spectral libraries}
These methods assume the availability of a spectral library, which is an overcomplete collection of multiple signatures for each EM that represent the possible variability for that EM.
The simplest method that uses spectral libraries is Multiple Endmember Spectral Mixture Analysis (MESMA) \cite{roberts_mapping_1998}. MESMA assumes an LMM in each pixel, and for every pixel, it finds the best-fitting EM signatures using least-squares. This is a combinatorial problem, so it can quickly become prohibitively large. Several variants exist, but they are only effective for small spectral libraries. As a way to reduce the computational burden of MESMA, sparse unmixing was proposed. Only one least-squares problem per pixel is solved, while enforcing abundance sparsity to limit the number of signatures that are actually contributing to the pixel. A well-known method for sparse unmixing is the Sparse Unmixing by variable Splitting and Augmented Lagrangian (SUnSAL) \cite{bioucas-dias_alternating_2010}, of which several variants exist \cite{iordache_total_2012}.

\subsubsection{Parametric physics-based models}
Physics-based models explicitly model the physical interaction of light with a material, described by a (relatively low) number of parameters. 
For example, the community standard for modeling topographic effects is Hapke's model \cite{hapke_theory_2012, heylen_review_2014}. Hapke's model uses the Single Scattering Albedo (SSA) of a material, in combination with incident and reflected angles, to determine the reflectance spectrum of a material. Physics-based models tend to be very complicated and very difficult, if not impossible, to invert. In the context of unmixing, this leads to extremely challenging and ill-posed optimization problems. 

\subsubsection{Parametric physically motivated models}
Since full physics-based models are too complicated, simpler models, which are not physics-based, but physically motivated, are often used \cite{borsoi_spectral_2021}. Starting from the LMM, several consequent generalizations have been proposed. 
Some approaches model the variability as a scaling of the EM signatures, which is especially suited for illumination- and topography-induced variability. The Scaled LMM (SLMM) includes a pixel-wise scaling factor. This simple model leads to a convex optimization problem. The Extended LMM (ELMM) is a generalization of the SLMM, and it includes a scaling factor for every EM in every pixel \cite{drumetz_blind_2016}. Next, the Generalized LMM (GLMM) incorporates wavelength-specific effects and includes additional scaling factors for each spectral band \cite{imbiriba_generalized_2018}. 
The variability can also be modeled as an additive perturbation, accounting for intrinsic variability and inter-class variability, i.e., variability due to the presence of unknown materials. The Perturbed LMM (PLMM) includes a pixel-wise additive perturbation on the EMs \cite{thouvenin_hyperspectral_2016}. The Augmented LMM (ALMM) combines the pixel-wise scaling factor from the SLMM with the additive perturbation of the PLMM \cite{hong_augmented_2019}.
Most of these models are non-convex and are solved using an alternating least squares (ALS) approach, which iteratively solves a series of convex least-squares sub-problems.

\subsubsection{Models jointly estimating EMs and abundances}
The performance of the models above depends heavily on the quality of the reference EMs. When they are provided a priori, we have no control over their quality. When pixels in an image are highly mixed or corrupted by noise, EMs extracted from the image by an EM extraction algorithm (EEA) are inaccurate. Because of this, it has been proposed to jointly estimate EMs and abundances. This allows the final EM estimates to vary from the initial estimates, which can help account for inaccurate initial EM estimates. Furthermore, if the EM estimates are allowed to vary pixel-wise, this approach is also effective in handling intrinsic variability.
Performing this joint estimation under spectral variability is based on nonnegative matrix factorization \cite{lee_learning_1999}. In \cite{drumetz_spectral_2020}, an extension of the ELMM is proposed in the robust ELMM (RELMM) using a quadratic volume regularization term for joint estimation under scaling variability. In \cite{yuan_projection-based_2015}, a library-based sparse NMF method is designed, which accounts for the difference in acquisition conditions between the library and the image by adaptively combining library spectra to approximate the true EMs in the image. The authors in \cite{uezato_hierarchical_2020} use a bundle-based approach and the specific problem geometry to estimate the abundances and EMs in each pixel. Many other approaches for joint estimation exist \cite{song_weighted_2022, sun_blind_2022}, but they do not incorporate spectral variability.


\subsubsection{Machine Learning methods}
Like in many fields of HSI, machine learning (ML) methods have been used to perform unmixing in the presence of spectral variability. In recent years, deep learning (DL) methods have become very popular, and numerous DL methods specifically designed for spectral variability have been proposed.
Many works use autoencoders (AEs) as a backbone. In \cite{chen_dsfc-ae_2024}, a new deep shared fully connected AE unmixing network was developed. 
Authors in \cite{shi_deep_2022} presented a variational AE-based model for spatial-spectral unmixing with EM variability, by linking the generated EMs to the probability distributions of EM bundles extracted from the hyperspectral image, and used adversarial learning to learn realistic EMs. In \cite{shi_probabilistic_2022}, the authors presented PGMSU: a two-stream network, including a variational AE for EM extraction and a fully connected network for abundance estimation. In \cite{zhang_spectral_2022}, a two-stream convolutional autoencoder, one for learning the abundances and one for learning variability coefficients, was used to explicitly account for variability. 

Other works incorporate physical models for handling variability. ReDSUNN \cite{borsoi_dynamical_2023} is a recurrent neural network (RNN) based model designed to handle both spatial and temporal variability, inspired by the GLMM. In \cite{cheng_hyperspectral_2023}, the authors designed a scaled-and-perturbed LMM (SPLMM) and used it in combination with a multi-stream feed-forward neural network. Authors in \cite{gao_proportional_2024} used a similar physical model, called the Proportional Perturbation Model (PPM). In \cite{gao_reversible_2024}, the authors presented RevNet: a CNN-based abundance estimation network combined with a reversible EM learning module, where the EMs are modeled by a parametrized probability distribution. Recently, the attention mechanism has also gained traction in unmixing under variability, e.g., \cite{su_multi-attention_2023} uses attention to discover global spatial features and to exploit redundancy in the spectral bands. 

A big downside of DL methods is their lack of interpretability, even though some efforts have been made to design (partially) interpretable networks (see, e.g., \cite{hong_endmember-guided_2022, zheng_blind_2024}). Training DL methods is also very expensive and unstable.

\subsection{Endmember extraction algorithms} \label{sec: endmember extraction}
EEAs can be used to perform \textit{blind unmixing}, where unmixing is performed without prior knowledge of the EM spectra. Note that we use blind unmixing to describe any unmixing process that does not involve prior knowledge of EM spectra. Joint estimation, as described in Sec. \ref{sec: unmixing under variability}, is one example of such a process, but we use blind unmixing as a broader term that also encompasses other algorithms.

In this paper, we will use two main categories of EEAs, and we assume throughout that the number of EMs is known a priori. The first category is \textit{pure pixel-based EEAs}, which assume the presence of pure pixels in the image. These algorithms select the EMs from the pixels in the image. Most available methods, such as the Pixel Purity Index \cite{boardman_mapping_1995} and N-FINDR \cite{winter_n-findr_1999}, are designed for data conforming to the LMM. On the other hand, Vertex Component Analysis (VCA) \cite{nascimento_vertex_2005} is also effective under variability, since it uses successive orthogonal projections, which can handle scaling variability.

Secondly, \textit{volume-based EEAs} try to identify a minimum-volume simplex that contains all data, and then select the vertices of this simplex as the EMs, even when no pixels occupy the vertex positions. Most methods either approximate the volume using determinant formulations (e.g., Sisal \cite{bioucas-dias_variable_2009}, RMVE \cite{fu_robust_2016}) or a quadratic term \cite{zhuang_regularization_2019}. When scaling variability is present, these methods fail, since the geometric simplex interpretation no longer holds and the data now occupy a polyhedral cone. This can be mitigated by the use of a projection step to project the data onto a simplex \cite{fu_nonnegative_2019}.

\subsection{Contributions}

Existing models accounting for spectral variability result in highly nonconvex and ill-posed optimization problems, which are difficult to solve. Furthermore, these models rely on simple first-order optimization algorithms, which do not leverage higher-order information and quickly get trapped in suboptimal solutions. To overcome these problems, we propose the \textbf{Two-Step Linear Mixing Model (2LMM)}, a new physically motivated model that balances computational tractability and model complexity. Our main contributions are as follows:
\begin{enumerate}
    \item The 2LMM achieves an \textbf{effective balance} between the mathematical ease of simple models like the LMM and SLMM, and the rich modeling capacity of complex models like the ELMM and GLMM. This results in a model that is mathematically tractable and retains a strong physical motivation. Moreover, the resulting unmixing problem is only mildly nonconvex. This mitigates common issues related to nonconvexity, such as over-reliance on the initial guess and the need for careful hyperparameter tuning.
    \item To efficiently solve this mildly nonconvex problem, we introduce an unmixing algorithm based on the BFGS algorithm, which \textbf{incorporates second-order derivative information}. To the authors' knowledge, this is the first time this is done in unmixing. This accelerates convergence and can enhance the accuracy of the solution. For more complex models, using second-order information is intractable. For simpler models, they have no added benefit, since these models lead to convex optimization problems that are already solved to global optimality by first-order methods. This makes the 2LMM uniquely suitable for using a BFGS algorithm.
    \item The proposed approach is validated with experiments on simulated data, along with real data: a \textbf{new benchmark dataset} with spectral variability and ground truth \cite{haijen_benchmark_2025}, and a dataset with topography-induced variability. The method is demonstrated to work in conjunction with EEAs in blind unmixing and highly mixed scenarios.
\end{enumerate}

\subsection{Outline}
The remainder of this article is structured as follows: in the next section, some related work and related mixing models are described. Section III is devoted to the description of the proposed model, along with the proposed optimization procedure. Experiments are conducted in section IV. Section V concludes the work.

\section{Related work} \label{sec: related work}
In this section, we first introduce the necessary notations. Then we expand on three physically motivated linear mixing models of interest, more precisely the LMM, SLMM, and ELMM. Then, we provide some background on EM extraction under variability.

\subsection{Notation}

We will denote by $K$ the number of EMs, by $N$ the number of pixels, and by $P$ the number of spectral bands. The abundance matrix is written as $\A \in \real^{K \times N}$, and the abundance vector of the $n$-th pixel is denoted by $\mathbf{a}_n, n = 1,2,\ldots, N$. EMs are denoted by the matrix $\E \in \real^{P\times K}$, and individual EMs by $\e_k, k=1,2,\ldots,K$. An image matrix is denoted by $\mathbf{X} \in \real^{P \times N}$, and a single pixel is written as $\x_n, n=1, 2, \ldots, N$. Let $\|\mathbf{v}\|_p$ denote the $p$-norm of a vector $\mathbf{v} \in \real^D$
and let $\|\mathbf{B}\|_{p, q}$ denote the $L_{p, q}$-norm of a matrix $\mathbf{B} \in \real^{M \times D}$.
In particular, let $\|\mathbf{B} \|_{2,2} := \|\mathbf{B} \|_F$ denote the Frobenius norm of a matrix. Furthermore, let $\mathrm{diag}(\mathbf{v})$ denote the diagonalization operator of a vector $\mathbf{v}$, such that $\mathrm{diag}(\mathbf{v}) \in \real^{D \times D}$ is a diagonal matrix with the elements of $\mathbf{v}$ on its diagonal. Finally, we will use the hat symbol $\hat{\cdot}$ to denote estimated quantities, to distinguish them from measured or ground truth quantities.

\subsection{Unmixing with the LMM} \label{sec: lmm}
The LMM assumes that every pixel can be written as a convex combination of EMs, which are the same across the entire image. Under the LMM, variability in the scene is only caused by EMs appearing in different abundances. The LMM can be written as:
\begin{equation} \label{eq: linear mixing model}
    \widehat{\x}_n = \sum_{k=1}^K \e_{k} a_{kn} =  \E \mathbf{a}_n
\end{equation}
where there are two constraints to be imposed on $\mathbf{a}_n$ in order for the abundances to satisfy the convex combination constraint, and to make them physically meaningful: the \textit{abundance non-negativity constraint} (ANC) $\mathbf{a}_n \geq 0$, and the \textit{abundance sum-to-one constraint} (ASC) 
$\sum_{k=1}^K a_{nk} = \mathbf{1}^\top \mathbf{a}_n = 1$.

Performing unmixing with the LMM can be done using a convex least-squares optimization approach. Denote by $\x_n$ the ground truth pixel. Then the optimization problem reads:
\begin{equation} \label{eq: fclsu}
    \begin{aligned}
    \min_{\mathbf{a}_n} &~ \frac{1}{2} \|{\x}_n - \E \mathbf{a}_n\|^2_2 \\
    \mathrm{s.t.} &~ \ba_n \geq 0 \quad \text{(ANC)} \\
    &~\mathbf{1}^\top \ba_n = 1 \quad \text{(ASC).}
    \end{aligned}
\end{equation}
This is a convex quadratic program with both equality and inequality constraints. Therefore, one of the many well-established methods for solving constrained quadratic programs can be used to solve it. A well-known method is the active set method \cite{nocedal_numerical_2006}. This method works by identifying a set of active constraints, which are the constraints that are currently being treated as equality constraints. The algorithm then solves a simpler subproblem with only equality constraints, and iteratively updates the active set until the optimal solution is found. In this work, we use the \texttt{lsqlin} method from the Mosek toolbox\footnote{
Available at \url{https://www.mosek.com/}.
}. We will refer to this method as \textbf{LMM}.

\subsection{Unmixing with the SLMM} \label{sec: slmm}
The SLMM assumes that EMs can change from pixel to pixel by means of a scaling, and this scaling is the same for all EMs within a pixel. In this way, the SLMM generalizes the LMM by introducing a pixel-wise scaling factor $s_{\x_n} > 0$:
\begin{equation}
    \widehat{\x}_n = s_{\x_n} \E \ba_n.
\end{equation}
Unmixing with the SLMM is commonly done by a non-negative least-squares approach, by dropping the ASC in problem (\ref{eq: fclsu}). This method is called the (partially) constrained least-squares unmixing (CLSU). The non-normalized abundances $\A_\s$, which are defined columnwise as $\mathbf{a}_{\s, n} = s_{\x_n}\ba_n$, can be found analytically using the normal equations, followed by a projection onto the nonnegative orthant:
\begin{equation}
    \A_\s = \max \{ (\E^\top \E)^{-1}\E^\top\X, ~0\}
\end{equation}
where the $\max$ operator is applied elementwise. Performing this operation directly is very ill-conditioned and prone to errors. Therefore, software like \textsc{Matlab} uses better conditioned operations, such as QR factorization, to indirectly solve the normal equations. The abundances are then obtained via a post-processing normalization step:
\begin{equation}\label{eq: normalization}
            s_{\x_n} = \sum_{k=1}^K a_{\s, nk}, \quad \ba_{n} = \frac{\ba_{\s, n}}{s_{\x_n}}.
\end{equation}
We will refer to this method as \textbf{SLMM}.

\subsection{Unmixing with the ELMM}
Since spectral variability is often material-specific (e.g., in the case of topography-induced variability in Hapke's model), the SLMM fails to accurately model many real-world scenes. Rather than a single scalar, the ELMM introduces a scaling vector in every pixel, allowing each EM to be scaled differently in every pixel, which facilitates modeling more complex and material-specific variability \cite{drumetz_blind_2016}.  Define a pixel scaling vector $\s_n \in \real^K$ for every pixel, then the ELMM reads:
\begin{equation}
    \widehat{\x}_n = \E_n \mathrm{diag}(\s_n) \ba_n.
\end{equation}
The pixel-wise EMs $\E_n$ are slightly perturbed versions of the reference EMs $\E$. Performing unmixing with the ELMM is done using a regularized version of the least-squares cost function, which iteratively updates the abundances and scaling factors. For the update of the abundances $\A$, an iterative algorithm is used, while the updates for the scaling factors $\bS$ and for the pixel-wise EMs $\E_n$ can be done analytically. The algorithm is initialized with the abundance estimates obtained from the SLMM to improve performance. The ELMM is highly non-convex and it requires careful tuning of the regularization parameters to achieve the best possible performance. As will be shown in the experiments, it is also very dependent on the initialization.

\subsection{Endmember extraction under variability}

Most EEAs are designed to work with LMM-conforming data. In this case, the data lies on a simplex, so the problem has a simple geometric structure. When scaling variability is present, the data no longer lie on a simplex, but instead occupy a polyhedral cone \cite{drumetz_spectral_2020}. Since the origin is also an extreme point of this data, this can cause spurious EMs. To mitigate this, we use VCA as a representative of the pure pixel-based class of EEAs, since it is inherently capable of addressing scaling variability. VCA uses a perspective projection $\mathbf{proj}(\x)$, which projects the conical data onto a simplex \cite{fu_identifiability_2018}:
\begin{equation} \label{eq: perspective projection}
            \mathbf{proj}: \real^P \rightarrow \real^P: \x \mapsto\mathbf{proj}(\x) = \frac{\x}{\x^\top \mathbf{v}}
\end{equation}
where $\mathbf{v} \in \real^P$ is a vector such that $\x^\top \mathbf{v} \neq 0$ for all $\x$ in $\mathbf{X}$.

Despite this indifference to scaling, VCA still requires pure pixels in the image. This is formally known as the \textit{separability assumption}, and it requires pixels at the simplex vertices. Volume-based methods do not require this restrictive assumption, and as a result, they also work on images without pure pixels. Finding necessary conditions under which volume-based EEAs are guaranteed to work is still an active area of research. A sufficient condition is known, and is called the sufficiently scattered condition \cite{fu_nonnegative_2019}:
\begin{definition}
    A nonnegative matrix $\A \in \real^{K \times N}$ is said to be \textbf{sufficiently scattered} if and only if (1) $\mathcal{C} \subseteq \mathrm{cone}(\A)$, where $\mathrm{cone}(\A)$ is the polyhedral cone generated by the columns of $\A$ and $\mathcal{C}$ is the second-order cone defined as 
    \begin{equation}
        \mathcal{C} = \{ \x \in \real^K \mid \x^\top \mathbf{1} \geq \sqrt{K -1}\|\x\|_2\}, 
    \end{equation}
    and (2) $\mathrm{cone}(\A) \subseteq \mathrm{cone}(\mathbf{Q})$ does not hold for any orthonormal $\mathbf{Q}$ except the permutation matrices.
\end{definition}

\begin{figure}
    \centering
    \includegraphics[width=0.6\linewidth]{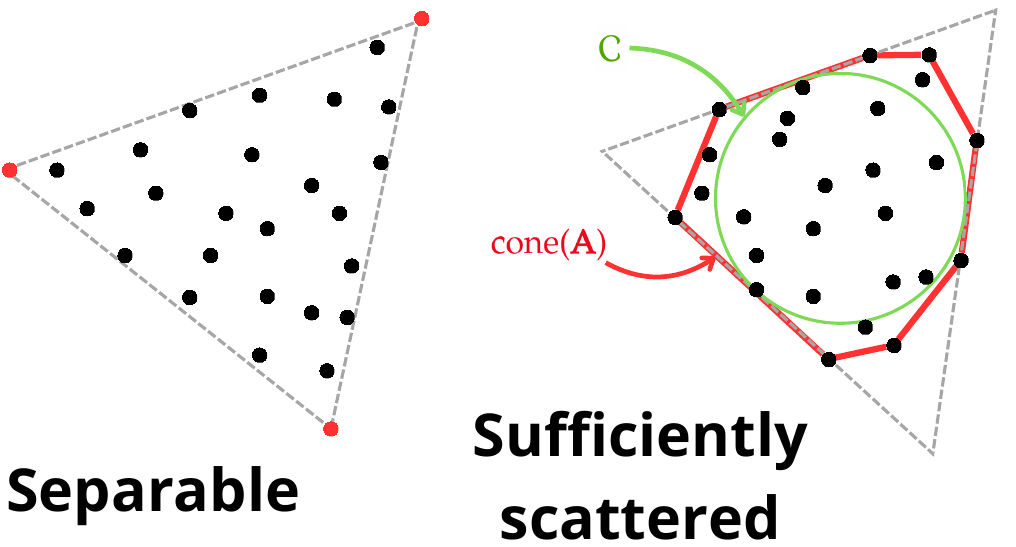}
    \caption{An illustration of the separability (left) and sufficiently scattered (right) condition.}
    \label{fig: separable and sufficiently scattered}
\end{figure}

The two definitions are illustrated in Fig. \ref{fig: separable and sufficiently scattered}. The separability assumption requires pure pixels, indicated by the red dots. The sufficiently scattered condition requires the polyhedral cone generated by $\A$ (red contour) completely contain the second-order cone inscribed in the simplex, $\mathcal{C}$ (green contour). This means that there must be pixels on all faces of the simplex.


Volume-based methods cannot be used directly on conical data, since the origin is also an extreme point of this data. Therefore, we use a perspective projection on the data prior to EM extraction. In this work, we use Sisal as volume-based EEA \cite{bioucas-dias_variable_2009}, which uses the log determinant of the dimension-reduced EM matrix as a volume proxy, and then minimizes this volume while ensuring a good fit to the data.

When the sufficiently scattered condition is satisfied, volume-based EEAs will recover the true EMs, up to a scaling factor:
\begin{theorem} \label{thm: EM extraction}
    If both ground truth factors $\Tilde{\E}$ and $\Tilde{\A}$ are sufficiently scattered, then any extracted solution $\widehat{\E}$ must satisfy $\widehat{\E} = \Tilde{\E} \bm{\Pi} \mathbf{D}$, where $\bm{\Pi}$ is a permutation matrix and $\mathbf{D}$ is a full-rank diagonal matrix.
\end{theorem}
A similar result exists for pure pixel-based methods, when the separability condition is satisfied \cite{fu_identifiability_2018, fu_nonnegative_2019}.

\section{Proposed model (2LMM)}
\subsection{Model description} 

\begin{figure}
    \centering
    \includegraphics[width=\linewidth]{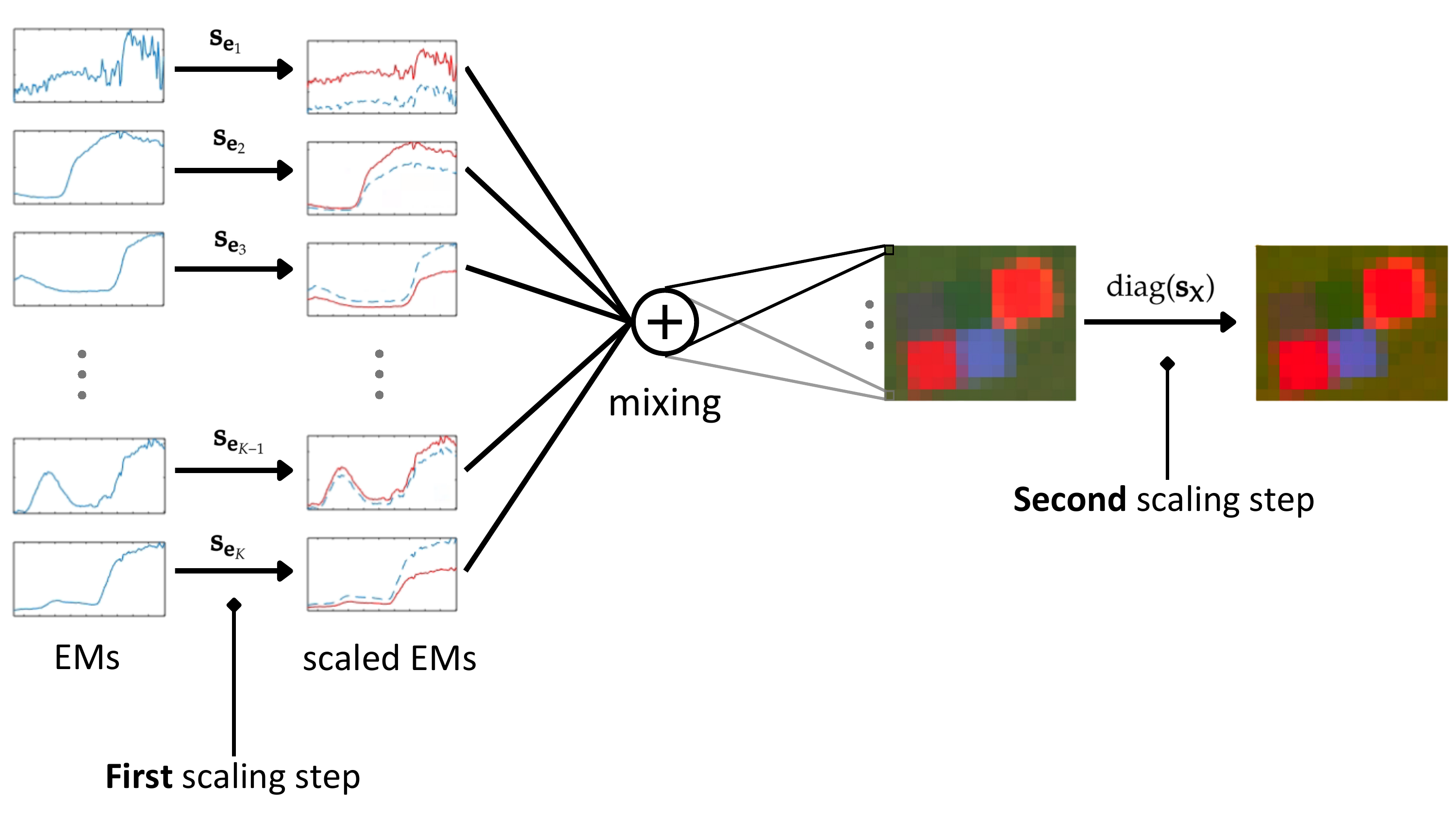}
    \caption{A graphical representation of the 2LMM model assumption. \textbf{First scaling step}: The reference EMs (blue lines) are scaled independently (red lines). \textbf{Mixing}: The scaled EMs are mixed to form the unscaled pixels in the image. \textbf{Second scaling step}: Each pixel is scaled independently to form the final image.}
    \label{fig: 2lmm concept}
\end{figure}

In this section, we describe in detail the \textbf{2LMM}, which bridges the gap between the simple SLMM and the rich, but complicated ELMM. Using the reference EMs $\E$, the model is constructed as follows. As a first scaling step, the EMs are scaled independently from one another, but in the same way across the entire image. Then, the EMs are linearly combined to form unscaled pixels. The second scaling step then consists of scaling each mixed pixel independently. See Fig. \ref{fig: 2lmm concept} for a conceptual representation of the 2LMM mixing process. The 2LMM can still model material-specific variability, but in a more constrained way than the ELMM.

\subsection{Motivation}

The motivation for the new model is two-fold. The first argument is mathematical, since the 2LMM is a moderately complex model and thus allows for the use of second-order optimization. The second argument is physical, since the two scaling steps have a distinct physical meaning.

\subsubsection{Moderate complexity}

\begin{figure}
    \centering
    \includegraphics[width=\linewidth]{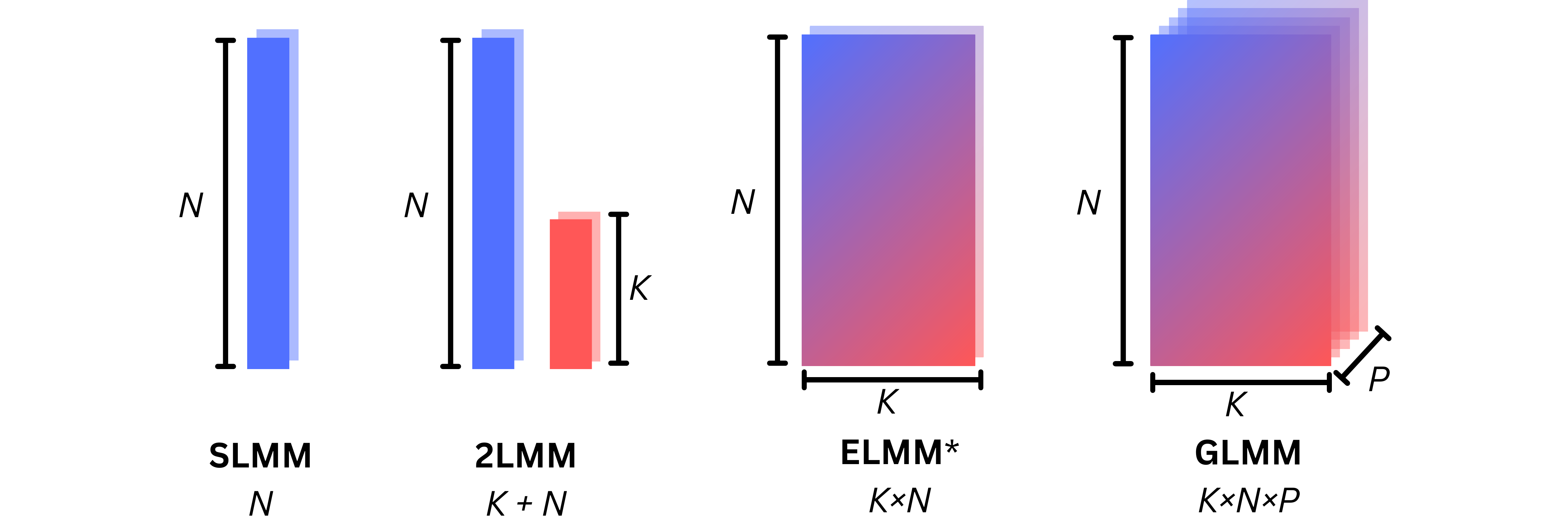}
    \caption{A graphical representation of the variability parameter count of several physics-inspired models. (*) the EM perturbation factors of the ELMM are not included since they contribute very little to model complexity.}
    \label{fig: parameters}
\end{figure}

First, the 2LMM balances the computational ease of the SLMM and the model complexity of the ELMM. To support this, we show the number of parameters each model has to estimate. First of all, all methods estimate the $KN$ abundance fractions. The SLMM has a total of 1 scaling factor per pixel, so a total of $N$ scaling factors. The 2LMM has $K$ additional scaling factors (in the EM scaling vector) for a total of $K + N$. The ELMM has a scaling factor in every pixel, for every EM, so this is a total of $KN$ scaling factors. Furthermore, the ELMM has EM perturbation factors for every pixel, which are a total of $KNP$ scalars. However, since these perturbation factors are heavily regularized, their effect on model complexity is limited. For illustrative purposes, we also include the GLMM \cite{imbiriba_generalized_2018}, which has wavelength-dependent scaling factors, totaling $KNP$ factors for the entire image. This is graphically represented in Fig. \ref{fig: parameters}. This moderate complexity results in an optimization problem that is only mildly non-convex, and therefore it does not suffer from numerical instabilities, sensitivity to the initial guess, or the need for careful hyperparameter tuning. At the same time, the 2LMM retains the physical modeling capacity of more complex models.
This moderate complexity also makes the 2LMM uniquely suitable for using second-order optimization methods. They can reduce the number of iterations required to reach convergence, and they can help in avoiding bad local minima. For more complex models, using second-order information is intractable due to the large memory requirements. For simpler models, second-order methods have no added benefit, since these models lead to convex optimization problems that are easily solved to global optimality by first-order or analytical methods. 

\subsubsection{Physically motivated}

Secondly, the two scaling steps have clear physical motivations. The first scaling step serves to correct any scaling discrepancies resulting from the provided EMs. If the EMs are obtained from a spectral library, the acquisition conditions used to generate the library might differ significantly, e.g., due to the use of a different sensor. On the other hand, if the EMs were obtained by an EEA, Thm. \ref{thm: EM extraction} states that this can introduce EM-specific scaling factors. Both of these cases benefit from the first scaling step, which is uniquely suitable to model their effects.
The second scaling step then further corrects for position-dependent effects, such as topography-induced variability or local illumination differences. 


\subsection{Mathematical formulation}

Let $\s_\E \in \real^K$ be the EM scaling vector representing the first scaling step, and $s_{\x_n}$ a pixel-dependent scaling factor representing the second scaling step for the $n$-th pixel. Then the $n$-th pixel in the 2LMM is given by:
\begin{equation}
    \widehat{\x}_n = \E\mathrm{diag}(\s_\E)\ba_n s_{\x_n}.
\end{equation}
We can combine this for all pixels. Denote the pixel scaling vector by
{\(
\s_\X = [s_{\x_1}~s_{\x_2}~\cdots~s_{\x_N}]^\top
\)}
and the estimated image matrix by $\widehat{\X}$. Then the 2LMM at the image level is given by:

\begin{equation} \label{eq: 2lmm cost function}
    \widehat{\X} = \E\mathrm{diag}(\s_\E)\A\mathrm{diag}(\s_\X).
\end{equation}

For performing unmixing, we use a nonlinear and non-convex constrained least-squares problem:
\begin{equation}
\begin{aligned}
    \min_{\A, \s_\E, \s_\X} &~\|{\X} - \E \diag(\s_\E)\A\diag(\s_\X)\|_F^2 \\
    \text{s.t.} &~ 0 \leq \A \leq 1, \mathbf{1}^\top \A = \mathbf{1} \\
    &~ \underline{S} \leq \s_\E \leq \overline{S}, \quad \underline{S} \leq \s_\X \leq \overline{S}
\end{aligned}
\end{equation}
where the box bounds $\underline{S}, \overline{S} > 0$ can be used to constrain the scaling variables to a user-specified interval. Naturally, $\underline{S} < \overline{S}$. The motivation for introducing these box bounds is both mathematical and physical. First, it makes the problem easier to solve mathematically, since it reduces the size of the search space, and therefore reduces the probability of finding a sub-optimal solution to the non-convex cost function. Second, it allows us to constrain the scaling factors to a physically meaningful range, since in many cases credible assumptions can be made about the magnitude of the scaling factors.


We simplify the problem by combining $\A$ and $\s_\X$ in the cost function (\ref{eq: 2lmm cost function}) into one matrix $\A_\s$, and subsequently dropping the ASC. This leads to the optimization problem:
\begin{equation} \label{eq: two scaling factor}
  (\mathrm{2LMM}):~ \begin{aligned}
        \min_{\A_\s, \s_\E} &~\|{\X} - \E \diag(\s_\E)\A_\s\|_F^2 \\
    \text{s.t.} &~ 0 \leq \A_\s \leq \overline{S},  \\
    &~ \underline{S} \leq \s_\E \leq \overline{S}
\end{aligned}
\end{equation}

We emphasize that combining the variables $\A$ and $\s_\X$ is a purely mathematical step, and it does not change the physical significance of the problem. As a result, $\A_\s$ is only a variable used during optimization, but not for further analysis. For analysis, we will always split $\A_\s$ into the actual abundances and pixel scaling factors -- which are the physical quantities of interest -- after optimization using the normalization procedure (\ref{eq: normalization}).

In our previous work \cite{haijen_two-step_2025}, we used an interior-point (IP) method for unmixing with the 2LMM. However, this algorithm is very memory-intensive, and as a result, it has a large computational cost for mid-sized images, and it is intractable for large images. Furthermore, we used a general-purpose solver, Ipopt, for solving the problem. The general nature of this solver makes it less efficient for the specific use case of nonlinear least-squares. Therefore, in what follows, we describe a new method for solving the 2LMM, which uses an alternating least-squares (ALS) approach, combined with the limited memory BFGS (L-BFGS) algorithm \cite{liu_limited_1989}. We expect this algorithm to be faster, albeit at the cost of some accuracy.

\subsubsection{Alternating least-squares for the 2LMM}

ALS is a popular approach for designing unmixing algorithms \cite{drumetz_blind_2016, imbiriba_generalized_2018, thouvenin_hyperspectral_2016}. We apply this approach to the problem (\ref{eq: two scaling factor}). There are two optimization steps: one over $\A_\s$, and one over $\s_\E$.

\textit{$\A_\s$--update.} We use constrained least squares for this update. This is a convex optimization problem:
\begin{equation}
     \begin{aligned}
        \A_\s^{t+1} = \text{arg } \min_{\A_\s} & ~ \|\X-\E \diag(\s^t_\E) \A_\s\|_F^2 \\
        \text{s.t. } & ~ 0 \leq \A_\s \leq \overline{S}
    \end{aligned}
\end{equation}

This problem can be solved analytically by setting the matrix partial derivative $\frac{\partial J}{\partial \A_\s}$ to zero and rewriting for $\A_\s$:
\begin{equation}
\A^{+}_\s = \diag\left( \frac{[\mathbf{1}]}{[\s^t_\E]}\right)(\E^\top \E)^{-1}\E^\top \X    
\end{equation}
where $[\cdot] / [\cdot]$ denotes elementwise vector division. Finally, enforcing the constraints leads to the next iterate:
\begin{equation} \label{eq: als A iteration}
    \A_\s^{t+1} = \max \{0, \min \{\overline{S}, \A^+_\s\}\}
\end{equation}
where the $\max$ and $\min$ operators are applied elementwise.

\begin{figure*}[t]
\normalsize
\begin{equation}\label{eq: als s intermediate update}
    s_{\e_k}^+ = \frac{\sum_{n=1}^N a_{nk}^{t+1} \e_k^\top \x_n - \sum_{n=1}^N a^{t+1}_{nk} \e_k^\top \left( \sum_{i=1, i \neq k}^K \e_i s^{t+\lambda}_{\e_i} a^{t+1}_{ni}\right)}{\|\e_k\|_2^2 \sum_{n=1}^N (a_{nk}^{t+1})^2} \quad \text{where} \quad \lambda = \begin{cases}
        1 ~ \text{if} ~ i > k \\ 0 ~ \text{otherwise}
    \end{cases}
\end{equation}
\hrulefill
\vspace*{1pt}
\end{figure*}

\textbf{$\s_\E$--update} We can again derive the update rules for every $s_{\e_k}$ analytically. We set the partial derivative $\frac{\partial J}{\partial s_{\e_k}}$ to zero and re-write for $s_{\e_k}$. This gives the update rule (\ref{eq: als s intermediate update}). As a final step, we enforce the box bounds:
\begin{equation} \label{eq: als s update}
    s_{\e_k}^{t+1} =\max \{\underline{S}, \min \{\overline{S}, s_{\e_k}^+\}\}.
\end{equation}

\subsubsection{Concept of L-BFGS methods}
The L-BFGS algorithm is a limited-memory quasi-Newton method which updates the solution at time $t$, $\z^t$, using the update rule:
\begin{equation}
    \z^{t+1} = \z^t + \gamma^t \mathbf{p}_t,
\end{equation}
where $\gamma^t > 0$ is the step size, and $\mathbf{p}_t$ is the search direction. The search direction is the product of a matrix $\mathbf{H}_t$ and the gradient of the cost function $\nabla J(\z^t)$:
\begin{equation}
    \mathbf{p}_t = \mathbf{H}_t \nabla J(\z^t)
\end{equation}
The matrix $\mathbf{H}_t$ is an approximation to the inverse Hessian of the function, $(\nabla^2_{\z \z} f(\z^t))^{-1}$, and is updated in each iteration. This avoids the computation and inversion of the full Hessian, which is computationally prohibitive. To further reduce the memory requirements (computing and storing $\mathbf{H}_t$ requires $\mathcal{O}(D^2)$ memory when $\z \in \real^D$), the L-BFGS method does not explicitly compute $\mathbf{H}_t$, but immediately approximates the matrix-vector product $\mathbf{H}_t \nabla J(\z^t)$.
The step size $\gamma^t$ is usually found using a step-length selection algorithm such as backtracking (see \cite[Ch. 3.5]{nocedal_numerical_2006}).

L-BFGS allows to include approximate second-order information in the optimization procedure, which can significantly increase convergence speed compared to first-order methods and avoid getting stuck in poor local optima. It does so without the extensive compute and memory requirements of full second-order methods like IP methods. This behavior can be used to accelerate ALS algorithms, as we describe in the next paragraph.

\subsubsection{Accelerating ALS with L-BFGS}

Researchers have long tried to improve the convergence properties of ALS, and one tool that has been used for this is \textit{preconditioning}. This transforms the optimization problem into a different form, which is easier to solve. The authors in \cite{de_sterck_nonlinearly_2018} developed an algorithm for tensor decomposition based on ALS and L-BFGS, which replaces the standard L-BFGS search direction $\mathbf{H}_t\nabla J(\z^t)$ with a nonlinearly modified version $\widehat{\mathbf{H}}_t(\mathcal{P}(\z^t))$. Crucially, this quantity depends on the \textit{ALS direction}, which we call $\mathcal{P}(\z^t)$:
\begin{equation}
    \mathcal{P}(\z^t) = \z^{t+1} - \z^t,
\end{equation}
which is equivalent to performing one iteration of the ALS algorithm, and then subtracting the starting point. The algorithm uses $\mathcal{P}(\z^t)$ as the starting search direction, as opposed to the gradient direction in classical L-BFGS. This can be seen as a nonlinear change of variables, from $\nabla J(\z^t)$ to $\mathcal{P}(\z^t)$, hence this type of preconditioning is referred to as \textit{transformation preconditioning}. The ALS direction is then nonlinearly modified based on second-order information to obtain the final search direction $\widehat{\mathbf{H}}_t(\mathcal{P}(\z^t))$. This new step size incorporates second-order information, and it is what makes the algorithm second-order. Next, for step size selection, the authors of \cite{de_sterck_nonlinearly_2018} propose a modified backtracking procedure, which allows for a slight increase in the cost function if necessary, whereas other backtracking procedures require a monotone decrease. This improves speed and robustness. 

\subsubsection{Implementation}
The algorithm is implemented in \textsc{Matlab}, as an extension to the code written by \cite{de_sterck_nonlinearly_2018}. The full conceptual algorithm is given in Algorithm \ref{alg:bfgs 2lmm}. Note that the box constraints are enforced during the computation of $\z^{t+1}$, and not during the step size selection. This means that iterates can (temporarily) leave the feasible set. However, we observed in our experiments that this did not have a negative impact on the result. We implemented a new termination condition, which depends on the rate of change of both $\A_\s$ and $\s_\E$ (see lines 8--10 of Algorithm \ref{alg:bfgs 2lmm}). We also implemented an efficient and analytical gradient calculation algorithm, which assists the L-BFGS algorithm during optimization. 
The output of the algorithm is the non-normalized abundance matrix $\A_\s$ and the scaling vector $\s_\E$. We then recover the abundances and pixel scaling factors using the normalization step (\ref{eq: normalization}). We will refer to this algorithm as \textbf{2LMM}.

\begin{algorithm}
\caption{2LMM unmixing with L-BFGS}\label{alg:bfgs 2lmm}
\SetKwInOut{Input}{Input}
\SetKwInOut{Output}{Output}

\Input{Image matrix $\X$; EM matrix $\E$; \\Initial iterate $\z^0 = [\A_\s^0; \s_\E^0]$; \\ Scaling bounds $\underline{S}, \overline{S}$ and thresholds $\epsilon_\A, \epsilon_\s$}
\Output{Non-normalized abundances $\A^\star_\s$; \\ EM scaling factors $\s_\E^\star$}
\For{$t = 1, 2, \ldots$}
{
    Perform one ALS iteration step according to Eqs. (\ref{eq: als A iteration}) and (\ref{eq: als s update}) and call the new iterate $\z^{+}$
    
    Define the nonlinear preconditioner $\mathcal{P}(\z^t) = \z^{+} - \z^t$
    
    $\mathbf{p}_t \leftarrow - \widehat{\mathbf{H}}_t(\mathcal{P}(\z^t))$
    
    (Modified backtracking) Find a sufficiently large $\gamma^t > 0$ such that
    
    $J(\z^t + \gamma^t \mathbf{p}_t) \leq (1 + e^{-t}) J(\z^t)$
    
    $\z^{t+1} \leftarrow \z^t + \gamma^t \mathbf{p}_t$
    
    \If {$\frac{\|\A_\s^{t-1} - \A_\s^t\|}{\|\A_\s^{t-1}\|} \leq \epsilon_\A ~~ \mathrm{and} ~~ \frac{\|\s_\E^{t-1} - \s_\E^t\|}{\|\s_\E^{t-1}\|} \leq \epsilon_\s$}
    {
        \KwRet{$\A_\s^\star = \A_\s^{t} ~~ \mathrm{and} ~~ \s_\E^\star = \s_\E^{t}$}
    }
    $t \leftarrow t + 1$
}
\end{algorithm}

\subsubsection{Convergence}

There is no formal convergence analysis for the nonlinearly preconditioned L-BFGS algorithm with the modified backtracking procedure. However, numerical experiments \cite{de_sterck_nonlinearly_2018} have demonstrated good convergence behavior, and a convergence proof for classical L-BFGS shows linear convergence under mild assumptions \cite{liu_limited_1989}.

\section{Experiments}
In this section, we validate and compare the newly proposed method to several state-of-the-art methods, including model-based and DL approaches, and to our earlier IP implementation of the 2LMM. Throughout these experiments, the results will be validated using the following metrics. Let $\x_n$ and $\hat{\x}_n$ denote real (measured) and estimated pixels, respectively, for all $n=1,2,\ldots, N$. The reconstruction Root Mean Square Error (RMSE) is defined as:
\begin{equation}
    \mathrm{RMSE}_\X = \sqrt{\frac{1}{NP} \sum_{n=1}^N \|\x_n - \hat{\x}_n\|_2^2}.
\end{equation}
Let $\mathbf{a}_n$ denote the actual abundance, and let $\hat{\mathbf{a}}_n$ denote the estimated abundance vector. To validate the performance of the abundance estimation, we define the abundance RMSE as:
\begin{equation}
    \mathrm{RMSE}_\A = \sqrt{\frac{1}{KN} \|\mathbf{a}_n - \hat{\mathbf{a}}_n\|_2^2}.
\end{equation}

We use several other physics-inspired and deep learning-based methods to compare with 2LMM: (1) \textbf{LMM} and (2) \textbf{SLMM}, solved using the algorithms described in Sec. \ref{sec: related work}, (3) \textbf{ELMM} \cite{drumetz_blind_2016}, (4) \textbf{GLMM} \cite{imbiriba_generalized_2018}, (5) \textbf{RevNet} \cite{gao_reversible_2024}, (6) \textbf{ReDSUNN} \cite{borsoi_dynamical_2023}, (7) \textbf{PGMSU} \cite{shi_probabilistic_2022}, and (8) \textbf{IP-2LMM}, our previously proposed unmixing algorithm using the 2LMM and an IP algorithm \cite{haijen_two-step_2025}. 

The ELMM method is initialized with SLMM. To compare the ELMM without this initialization, we include an ELMM version that is initialized with uniform abundances, and call this the cold-started version, ELMM(CS), compared to the warm-started version, ELMM(WS). 

To assess the impact of the L-BFGS algorithm, we also use a version of the 2LMM with a conventional alternating least squares (ALS), i.e., using the update rules (\ref{eq: als s update}) and (\ref{eq: als A iteration}) alternatively until convergence. We call this method ALS-2LMM. All experiments were run on a desktop computer with a 32-core Intel i9 CPU with a 3-level cache and 64 GiB RAM (DIMM).

\subsection{Synthetic data}

\begin{figure}[t]
    \centering
    \includegraphics[width=0.9\linewidth]{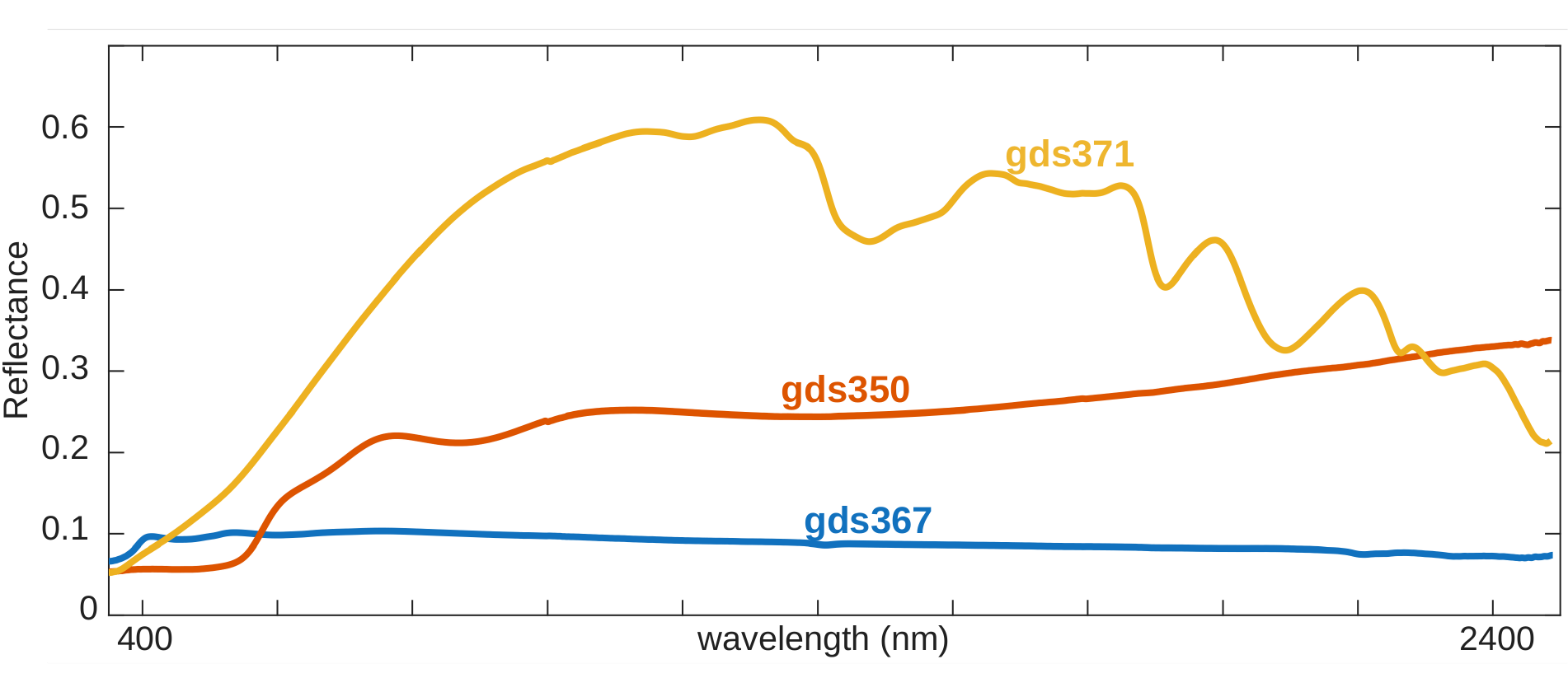}
    \caption{The EMs used for generating the synthetic data: asphalt (gds367), brick (gds350), and cardboard (gds371).}
    \label{fig: usgs ems}
\end{figure}

\begin{figure}[t]
    \centering
    \includegraphics[width=\linewidth]{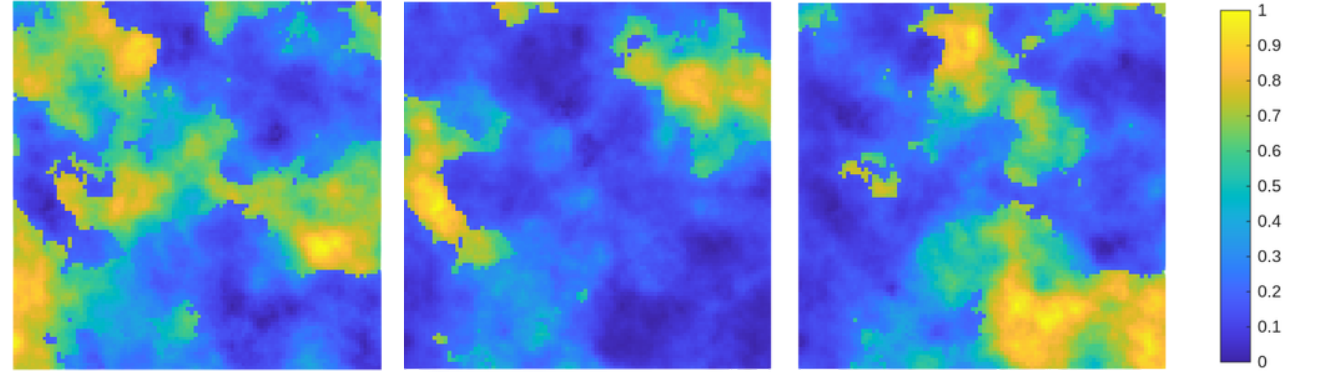}
    \caption{Abundance maps generated with Gaussian Random Fields for a synthetic image with three EMs.}
    \label{fig: abundance grfs}
\end{figure}

We select three EMs (asphalt (gds367), brick (gds350), and cardboard (gds371)) from the United States Geological Survey (USGS) spectral library \cite{kokaly_usgs_2017}, which contain 2152 spectral bands from the visible to the short-wave infrared range (200 nm to 2,500 nm).  We select 224 equidistant bands for each EM. We call these the reference EMs $\E_0$. Their reflectance is shown in Fig. \ref{fig: usgs ems}. We generate synthetic abundance maps based on Gaussian Random Fields (GRFs), which can be thought of as spatially correlated Gaussian randomness \cite{kozintsev_computations_1999}. GRFs are a popular choice for synthetic abundance generation\footnote{
We use the \textit{Hyperspectral Imagery Synthesis (EIAs) toolbox} for abundance generation, available at \url{https://www.ehu.eus/ccwintco/index.php/Hyperspectral_Imagery_Synthesis_tools_for_MATLAB}
}. We generate abundance maps using GRFs designed to comply with the ANC and ASC. Fig. \ref{fig: abundance grfs} shows an example of GRF abundance maps for an image with three EMs. We generate a $150 \times 150$ synthetic image. 

For introducing variability, we draw $N + K$ numbers from the uniform distribution $\mathcal{U}([\nicefrac{1}{3}; 3])$ and group them into vectors $\s_\E$ and $\s_\X$. Then we generate the $n$-th pixel as:
\begin{equation}
    \x_n = \E_0 \diag(\s_\E)\mathbf{a}_{\mathrm{gt}, n}s_{\x_n} + \bm{\epsilon}_n
\end{equation}
where $\bm{\epsilon}_n$ is normally distributed noise and $\mathbf{A}_\mathrm{gt}$ is the abundance matrix generated using GRFs. Even though this variability is generated according to the 2LMM model assumptions, more complex models such as the ELMM and GLMM possess sufficient modeling capacity to model any scene conforming to the 2LMM. Therefore, all three models (2LMM, ELMM, and GLMM) should -- in theory -- have no modeling error besides noise.

\subsubsection{Influence of the hyperparameters} 
\begin{figure}[t]
    \centering
    \includegraphics[width=\linewidth]{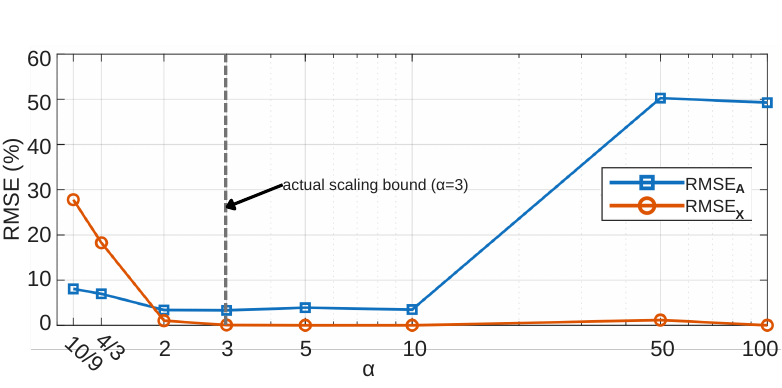}
    \caption{Unmixing results for a $150 \times 150$ synthetic image with varying scaling bounds for 2LMM. The bounds $[\underline{S}, \overline{S}]$ are given by $[\nicefrac{1}{\alpha}, \alpha]$.}
    \label{fig: bounds vs rmse}
\end{figure}

We add noise with an SNR of 40 dB. 2LMM is applied to this dataset. To examine the effect of the bounds on the resulting estimates, we vary the lower and upper bounds $\underline{S}$ and $\overline{S}$ and compare the performance. Fig. \ref{fig: bounds vs rmse} shows the reconstruction and abundance RMSE for 2LMM. If the scaling bounds are chosen too tight ($\alpha$ close to one), then the solution is considerably worse, since the optimal solution is not a part of the feasible set. When the scaling bounds are chosen to be approximately equal to the actual scaling bounds, then a (close to) optimal solution falls within the feasible set. When the bounds are considerably wider than the ground truth, the reconstruction is still very good, but the abundance estimation is poor. This is most likely due to the inclusion of many bad local minima in the feasible set. Since they might be located far from a global minimum, 2LMM fails to avoid them and gets trapped.

From a practical point of view, this means that the scaling bounds should be chosen sufficiently wide, such that the true scaling interval is within the chosen bounds. This ensures that the optimal solution lies within the feasible set. The experiment suggests that overestimating the scaling bounds by a factor of 2--3 does not impact performance. However, they must not be chosen excessively wide, i.e., orders of magnitude larger, since this leads to a poor abundance estimation.

\subsubsection{Ablation study}
We examine the effect of the L-BFGS mechanism on 2LMM unmixing. For this, we compare the performance of 2LMM to ALS-2LMM and IP-2LMM. For testing, we use a dataset that is generated in the same way as before, and we use $\{\underline{S}, \overline{S}\} = \left\{ \nicefrac{1}{5}, 5 \right\}$. The EMs are now extracted using VCA, which mimics a blind unmixing scenario, where EMs are not known beforehand. We compare both the performance and the computational cost: the runtime (in seconds), and the amount of memory allocated. 
Table \ref{tab: ablation} shows the results.

IP-2LMM performs the best, but has a high computational cost. Our newly proposed method is more than two orders of magnitude faster, and still performs quite well, with 7\% error. The advantage of the L-BFGS method is clearly demonstrated by the result of ALS-2LMM: without L-BFGS, 2LMM unmixing is both slower and less accurate. The additional memory cost compared to ALS-2LMM is limited: about 0.03 GiB.

\begin{table}[htb]
\centering
\caption{Comparison of performance and computational cost of three versions of 2LMM unmixing. The best results are highlighted in bold.}
\label{tab: ablation}
\begin{tabular}{|r|cccc|}
\hline
\multirow{2}{*}{\textbf{Method}} & \multicolumn{4}{c|}{\textbf{Metric}} \\
\cline{2-5}
& \textbf{RMSE}$_\A$ & \textbf{RMSE}$_\X$ & \textbf{Time (s)} & \textbf{Memory (GiB)} \\ \hline \hline
\textbf{IP-2LMM}  & \textbf{0.0238} & \textbf{5e-5}  & 157              & 40.23  \\
\textbf{ALS-2LMM} & 0.3819          & 0.0061         & 2.98             & \textbf{0.1137} \\
\textbf{2LMM}     & 0.0702          & \textbf{5e-5}  & \textbf{0.73}    & 0.1404 \\ \hline
\end{tabular}
\end{table}

\subsubsection{Performance under noise}

\begin{figure}
    \centering
    \includegraphics[width=0.8\linewidth]{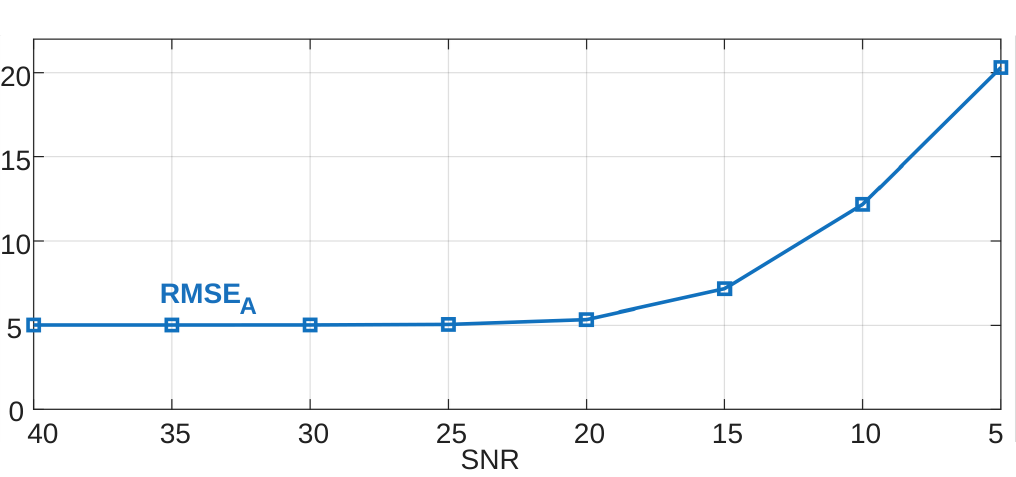}
    \caption{Abundance RMSEs (in \%) under various levels of noise for 2LMM on a synthetic image, which shows that 2LMM is robust to noise.}
    \label{fig: noise experiment}
\end{figure}

To demonstrate the robustness of the 2LMM to noise, we ran the experiment as described above for several noise levels. We used VCA to extract the EMs from the noiseless image. This way, we avoid that inaccuracies can be caused by VCA, and only study the noise robustness of 2LMM. The resulting abundance RMSE is shown in Fig. \ref{fig: noise experiment}. As long as the SNR is not extremely low, the result is only slightly affected by noise.

\subsubsection{Comparison to other methods}

\begin{table*}[t]
\centering
\caption{Experimental results on a synthetic image. The best results are highlighted in bold, and the second-best are underlined.}
\label{tab: 2lmm generated data}
\resizebox{\linewidth}{!}{%
\begin{tabular}{|l|cccccccccc|}
\hline
\multirow{2}{*}{\textbf{Metric}} & \multicolumn{10}{c|}{\textbf{Method}} \\
\cline{2-11}
& \textbf{LMM} & \textbf{SLMM} & \textbf{ELMM(WS)} & \textbf{ELMM(CS)} & \textbf{GLMM} & \textbf{RevNet} & \textbf{ReDSUNN} & \textbf{PGMSU} & \textbf{IP-2LMM} & \textbf{2LMM} \\
\hline \hline
\textbf{RMSE$_\mathbf{X}$} & 0.1201 & \textbf{5e-5} & \underline{1e-4} & 3e-4 & 8e-4 & 0.3313 & 0.1058 & 0.0948 & \textbf{5e-5} & \textbf{5e-5} \\
\textbf{RMSE$_\mathbf{A}$} & 0.2353 & 0.0578 & 0.0622 & 0.2415 & 0.1911 & 0.2373 & 0.3032 & 0.2980 & \textbf{0.0184} & \underline{0.0370} \\
\textbf{Time (s)} & 2.82 & \textbf{0.002} & 33.2 & 70.3 & 17.7 & 17.4 & 73.5 & 65.7 & 61.3 & \underline{0.39} \\
\hline
\end{tabular}
} 
\end{table*}

We follow the same procedure as before to generate the data, and use VCA to extract the EMs. For the 2LMM scaling bounds we chose $[\underline{S}, \overline{S}] = [\nicefrac{1}{5}, 5]$. The performance of the ELMM, GLMM, and deep learning-based methods can be improved with careful hyperparameter tuning. However, this is a highly nontrivial and data-dependent process, which is one of the key downsides that the 2LMM aims to address. Therefore, no hyperparameter tuning was carried out, and the default hyperparameters from the literature were used to provide a fair comparison.

The results are shown in Table \ref{tab: 2lmm generated data}. The 2LMM methods provide the most accurate results. While IP-2LMM has a slightly better abundance estimation, 2LMM achieved a considerably faster runtime. Lastly, the ELMM is highly dependent on a good initial guess; if the CLSU initialization is not provided, then the ELMM fails to perform well. This is not the case for the 2LMM, which is initialized using uniform abundances. In fact, additional experiments demonstrated that the 2LMM performance is largely insensitive to the initial guess, as long as it lies within the feasible set. More complex methods, such as the GLMM and Deep Learning methods, all perform considerably worse than 2LMM.

\subsection{Checkerboard dataset}
\begin{figure}[t]
    \centering
    \subfloat[Printed sample \label{fig:scene2}]{%
        \includegraphics[width=0.4\linewidth]{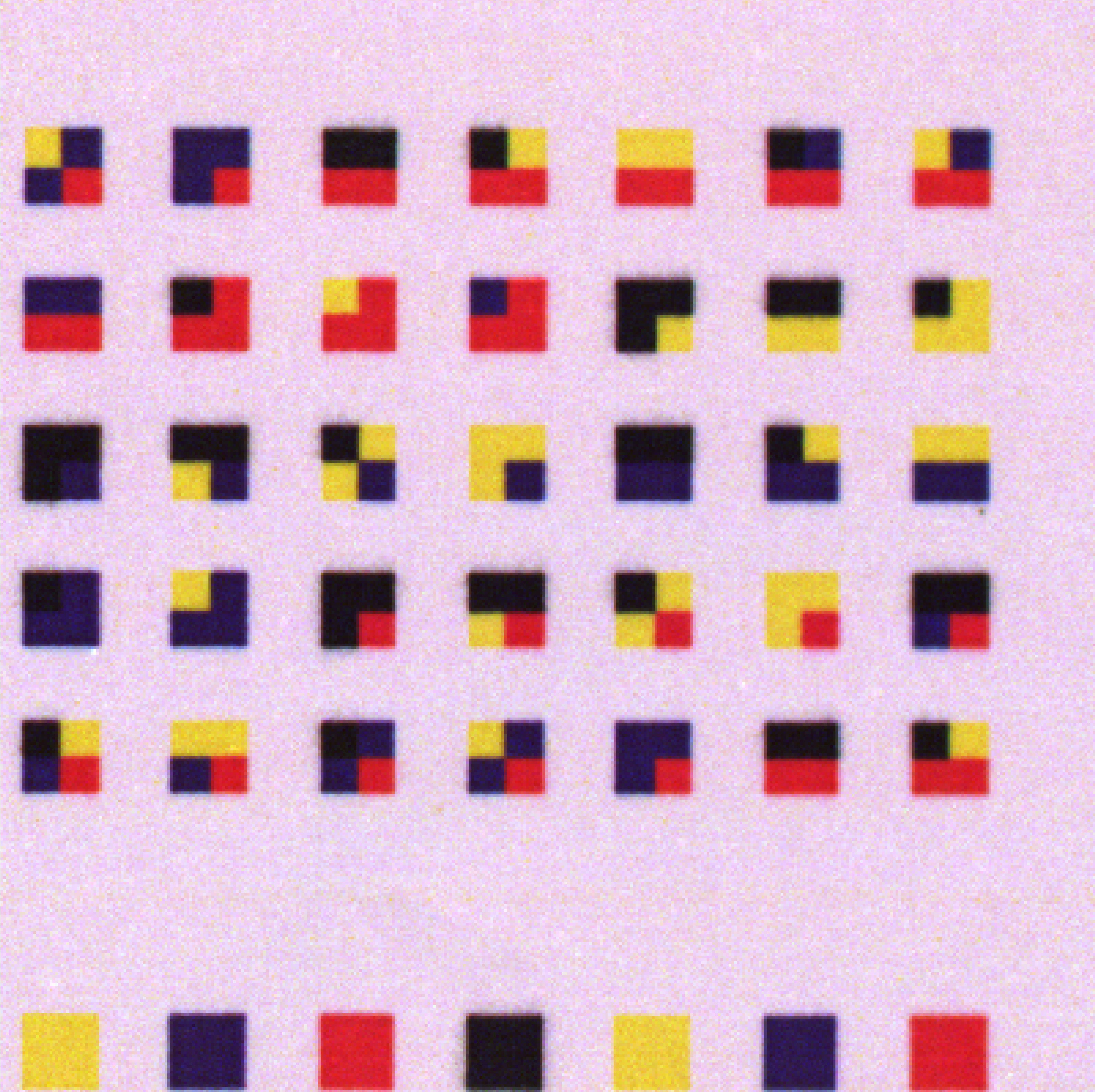}%
    }
    \hfill 
    \subfloat[Experimental setup \label{fig:experimentalsetup}]{%
        \includegraphics[width=0.55\linewidth]{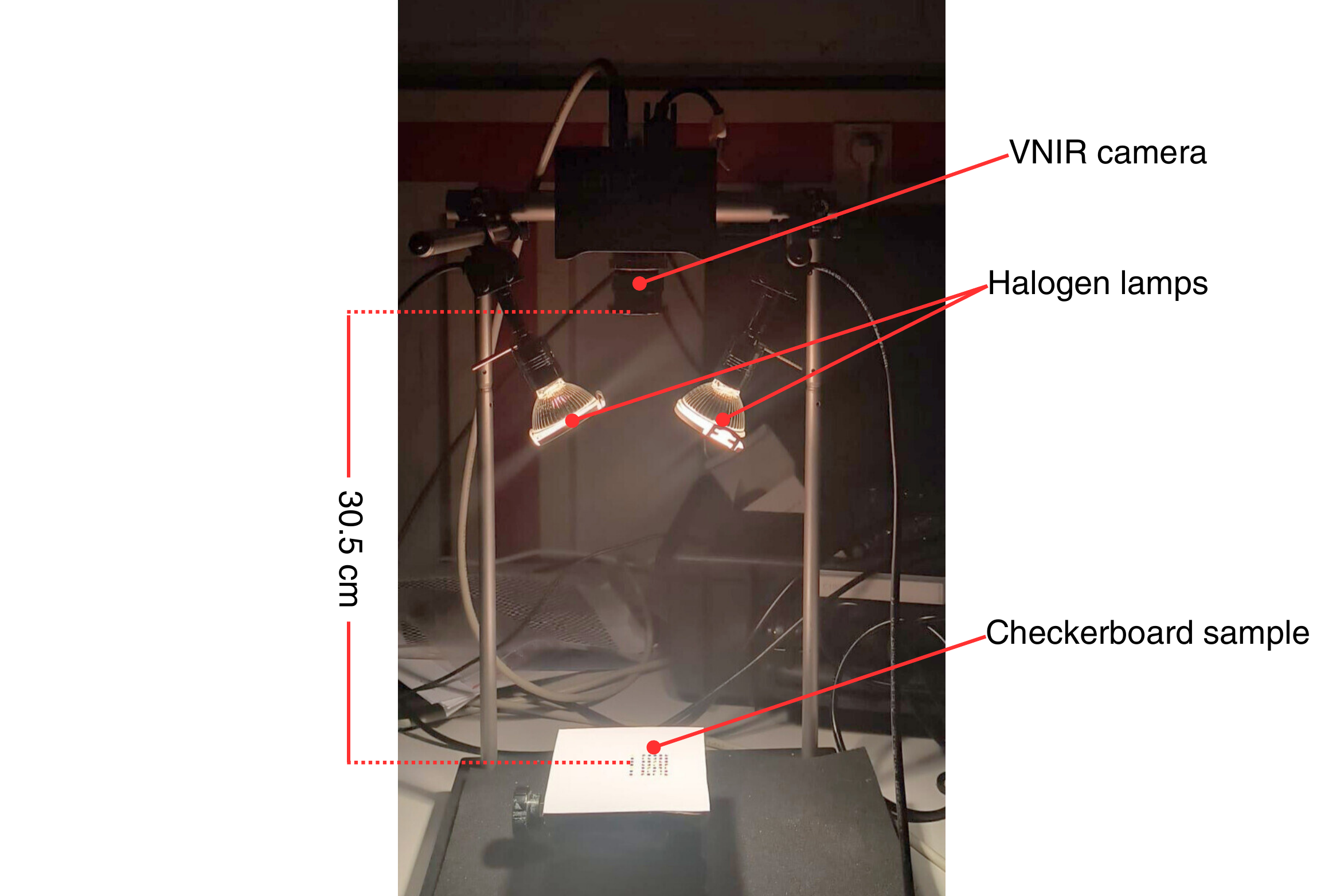}%
    }

    \caption{Two views of the experimental setup for the checkerboard dataset: (a) a printed sample, and (b) the overall setup. Each square represents one spectrum.} 
    \label{fig: combined setup}
\end{figure}

The checkerboard dataset is a dataset that has been recently developed by the authors and that contains spectral variability and abundance ground truth \cite{haijen_benchmark_2025}. The dataset consists of pure and mixed spectra made up of four colors, printed on paper. Fig. \ref{fig: combined setup} displays the sample and the experimental setup. The sample contains all possible combinations of the four colors with abundances $\{0; 0.25; 0.5; 0.75; 1\}$. The dataset was captured with the Imec Snapscan VNIR camera, which captures 150 spectral bands from 475 to 900 nm, at a spectral resolution of approximately 3 nm. Since the last few bands contained instrument artifacts, they were discarded, and the final data includes 130 bands.

\begin{figure}[t]
    \centering
    \includegraphics[width=\linewidth]{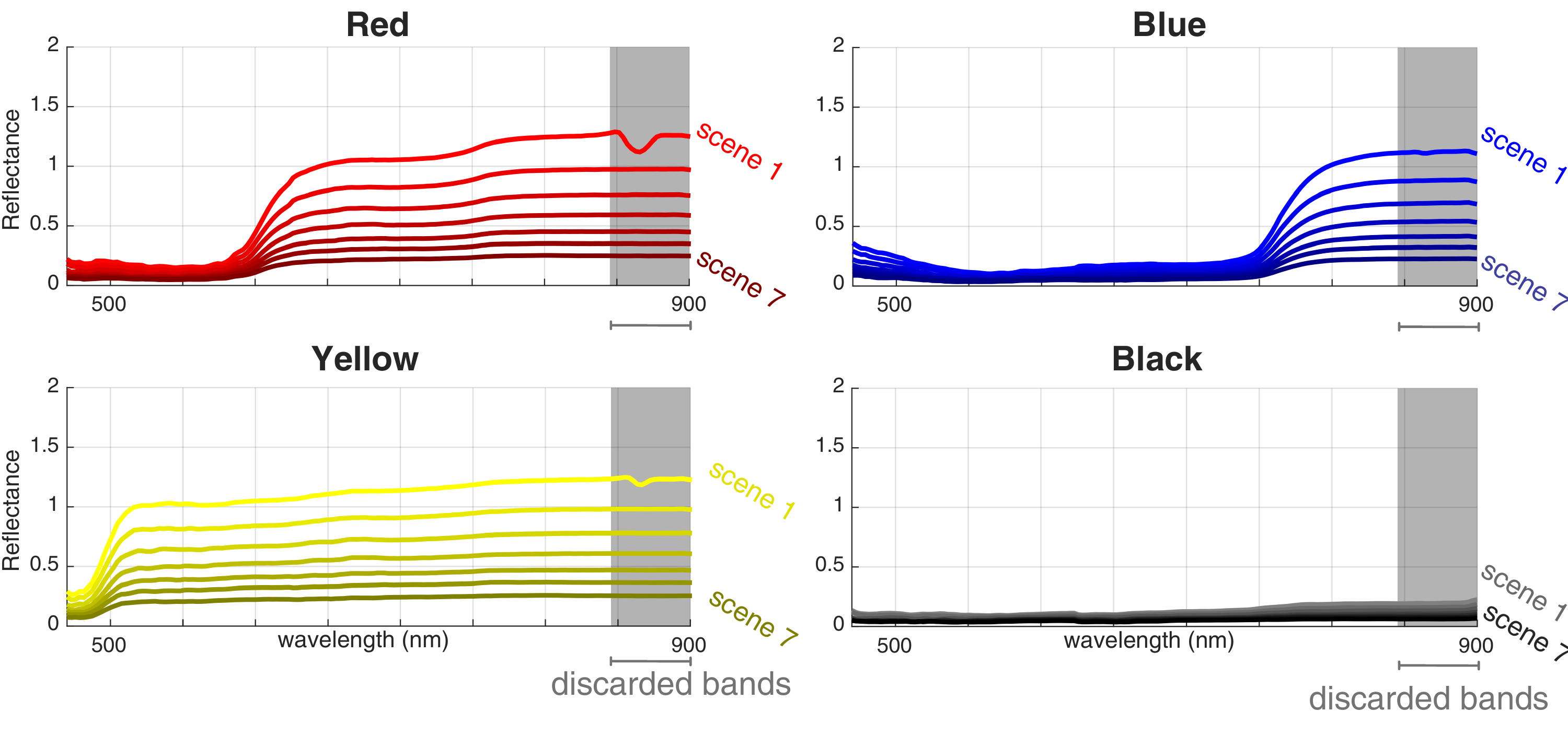}
    \caption{The four EMs for the seven acquisition scenes, manually extracted from every scene in the checkerboard dataset.}
    \label{fig: scaled ems}
\end{figure}

The light source is a pair of halogen lamps, which were placed at different distances from the sample to introduce spectral variability. In total, seven scenes were acquired for seven different settings of the halogen lamps. Moving the lamps causes an increase in illumination intensity, since the lamps are moved closer or further away from the sample. Secondly, the angle of incident light changes by moving the lamps. These two effects can be modeled quite well by scaling variability. As a result, the variability introduced in the dataset is almost exclusively scaling variability, which is illustrated in Fig. \ref{fig: scaled ems}, which shows the four EMs for the seven acquisition scenes. Additional variability, e.g., possible nonlinear effects and wavelength-dependent effects, are indistinguishable from the sensor-induced noise.

Each square represents one spectrum. Spectra were created by taking the mean of a $20\times20$ pixel region centered at the center of each square. Since black is a low-magnitude and featureless color, it is not suitable for unmixing validation. It will also lead to large errors when using the perspective projection (\ref{eq: perspective projection}), since the term $\mathbf{x}^\top\mathbf{v}$ in the denominator will be close to zero. Therefore, we exclude all pixels containing black. This results in a dataset containing 42 pure and 98 mixed spectra.

\begin{table*}[t]
\centering
\caption{Unmixing results for the checkerboard dataset using manually selected EMs and for a highly mixed scenario. The best results are highlighted in bold, and the second-best are underlined.}
\label{tab: checkerboard combined results}
\resizebox{\linewidth}{!}{%
\begin{tabular}{|r|r|ccccccccc|}
\hline
\multirow{2}{*}{\textbf{Experiment}} & \multirow{2}{*}{\textbf{Metric}} & \multicolumn{9}{c|}{\textbf{Method}} \\
\cline{3-11}
& & \textbf{LMM} & \textbf{SLMM} & \textbf{ELMM(WS)} & \textbf{GLMM} & \textbf{RevNet} & \textbf{ReDSUNN} & \textbf{PGMSU} & \textbf{IP-2LMM} & \textbf{2LMM} \\
\hline \hline
\multirow{3}{*}{\textbf{Manual}} & \textbf{RMSE$_\mathbf{X}$} & 0.1128 & 0.0101 & \textbf{2e-5} & 0.0010 & 0.2550 & 0.2864 & 0.0525 & 0.0086 & 0.0095 \\
& \textbf{RMSE$_\mathbf{A}$} & 0.3657 & 0.1997 & 0.2075 & 0.2832 & 0.3406 & 0.1977 & 0.3381 & \textbf{0.0562} & 0.0906 \\
& \textbf{Time (s)} & 0.06 & \textbf{0.001} & 10.8 & 3.29 & 3.05 & 3.44 & 20.9 & 0.32 & 0.14 \\
\hline
\multirow{3}{*}{\textbf{Highly mixed}} & \textbf{RMSE$_\mathbf{X}$} & 0.6758 & 0.0081 & \textbf{2e-5} & \underline{0.0018} & 0.5144 & 0.7543 & 0.0625 & 0.0080 & 0.0081 \\
& \textbf{RMSE$_\mathbf{A}$} & 0.4198 & 0.1288 & 0.1602 & 0.3417 & 0.2595 & 0.2222 & 0.2779 & \underline{0.1266} & \textbf{0.0683} \\
& \textbf{Time (s)} & 0.05 & \textbf{0.0003} & 12.1 & 4.02 & 2.93 & 1.65 & 21.0 & 0.26 & \underline{0.02} \\
\hline
\end{tabular}
} 
\end{table*}

\subsubsection{Manual selection of EMs}
For the first experiment, we manually select the EMs in order to introduce significant scale discrepancies. The red, blue, and yellow EMs were selected from the first, third, and seventh acquisition scenes, respectively. This situation directly mirrors the challenges posed by using EMs from spectral libraries. The 2LMM is specifically designed for such scenarios, and as expected, it outperforms all others. The first rows of Table \ref{tab: checkerboard combined results} show the unmixing results and highlight the superior performance of 2LMM. While some methods, like ELMM, achieved good reconstruction errors, this did not translate to accurate abundance estimations. The poor result for SLMM can be linked to its underlying assumption of similarly scaled EMs within a pixel. The ELMM and GLMM rely heavily on the SLMM for a good initialization, so they performed poorly as well. Deep learning methods also struggled to learn the variability.

\subsubsection{Highly mixed scenario}
In the second experiment, we mimic a highly mixed scenario by excluding the pure spectra from the dataset. The dataset now consists of 98 spectra, with a maximum abundance fraction of 75\%. Due to the absence of pure spectra, the separability assumption is violated. The sufficiently scattered condition is satisfied, since there are sufficient binary mixtures. Therefore, we now use Sisal with a perspective projection to obtain the reference EMs from the data. Since the data is noisy, the EM estimation is not perfect (the average Spectral Angle Distance (SAD) between manually selected and extracted EMs is approximately 2 degrees). The results are shown in the second part of Table \ref{tab: checkerboard combined results}. 2LMM is now the best-performing method, outperforming the state of the art by a large margin. It also outperforms IP-2LMM by 6\%. This demonstrates the suitability of 2LMM for blind unmixing under variability, in combination with a volume-based EEA.

\begin{figure}[t]
    \centering
    \includegraphics[width=\linewidth]{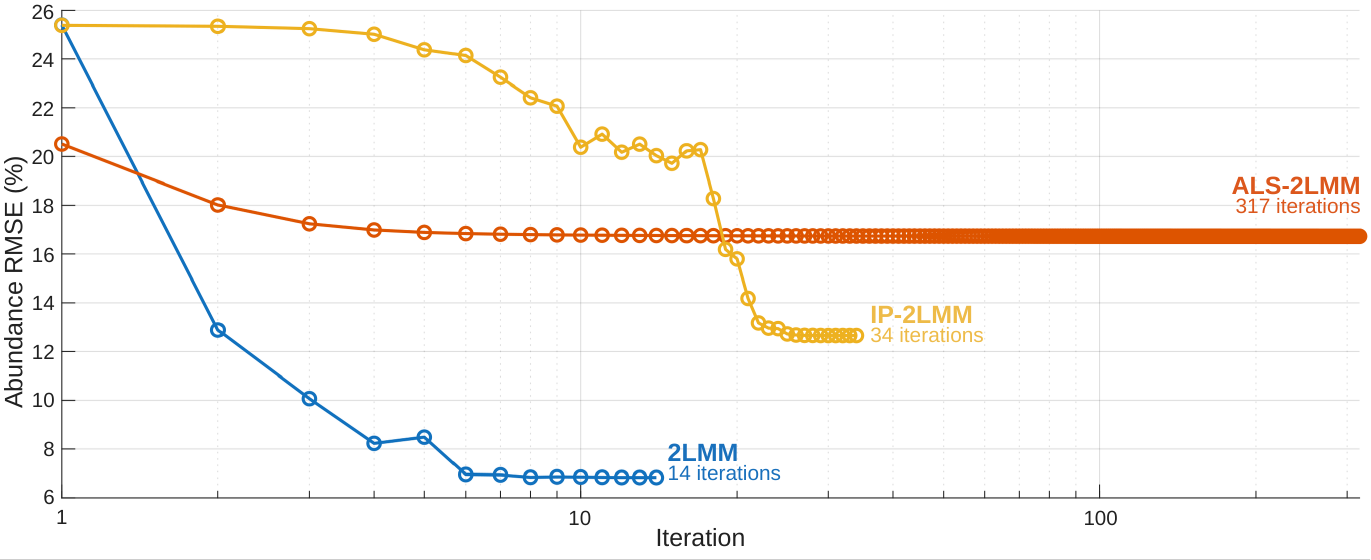}
    \caption{Abundance RMSE at each iteration for three 2LMM methods on the checkerboard dataset. The x-axis is on a logarithmic scale. Incorporating second-order information leads to faster convergence and more accurate solutions.}
    \label{fig: 2lmm convergence plot}
\end{figure}

This case can also be used to demonstrate the importance of using second-order information, both for the acceleration of convergence and for avoiding local minima. We ran the experiment as above for the three variants of 2LMM unmixing, and tracked the abundance error at every iteration in Fig. \ref{fig: 2lmm convergence plot}. Both IP-2LMM and 2LMM successfully avoid local minima and find a better optimum, while ALS-2LMM converges to a poor local minimum and is trapped. ALS-2LMM also requires substantially more iterations to reach convergence.

\subsection{Topography-induced variability}

In this experiment, we mimic the variability that is induced by changes in topography and combine it with realistic abundance maps to create a close to real-life unmixing scenario.

\subsubsection{Abundance maps and EMs}
\begin{figure}
    \centering
    \includegraphics[width=0.7\linewidth]{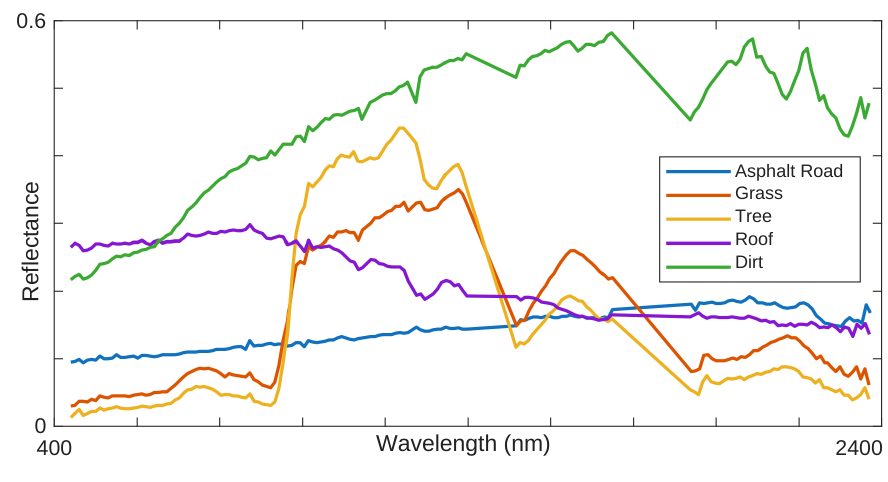}
    \caption{The five EMs from the Urban dataset}
    \label{fig: urban ems}
\end{figure}

For the abundance maps and EMs, we use the Urban dataset. This dataset consists of a $307 \times 307$ image, with 162 bands in the VNIR and SWIR (400 -- 2400 nm). We use the five EMs that are provided with the dataset, which are shown in Fig. \ref{fig: urban ems}. These EMs were manually extracted from the image. We designed two experiments, which we ran separately: a $100\times 100$ sub-image and the full $307 \times 307$ image. For the abundance maps, we use the ground truth abundances available with the dataset, which we call $\A_\mathrm{gt}$. They are shown in the top row of Fig. \ref{fig: urban abundances} for the $100\times100$ sub-image. 
We note that these ground truth abundances do not come from actual ground measurements, but were obtained using an LMM-based unmixing algorithm. Since the scene is flat, there is little variability, and a good correspondence between the actual ground cover and $\mathbf{A}_\mathrm{gt}$.

\subsubsection{Variability generation}

\begin{figure}
    \centering
    \includegraphics[width=0.8\linewidth]{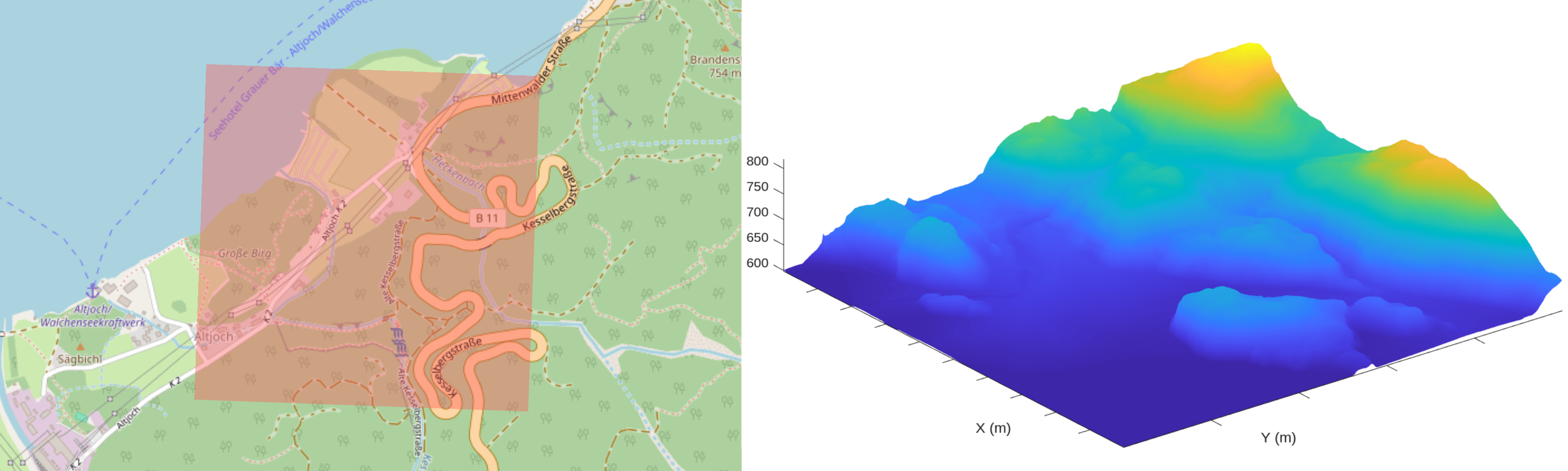}
    \caption{Location of the terrain (left) and the DSM (right) used for generating variability.}
    \label{fig: KochelAmSee}
\end{figure}

We use a real Digital Surface Model (DSM) from the village of Kochel Am See, Germany, which is located in the hilly Bavarian Prealps\footnote{
The data was retrieved from \url{https://geodaten.bayern.de/opengeodata/}.
}. It contains both flat and steep regions, representing a topographically diverse landscape. The location of the DSM, and the DSM itself, are shown in Fig. \ref{fig: KochelAmSee}. The DSM is used to derive the geometric parameters of the scene, including the slope of the terrain, the incidence angle of the light, and the emergence angle of the light. We use a 1 km$\times$1 km region of the DSM, and change the resolution depending on the experiment size: we sample the DSM at a resolution of 10 m for the $100\times100$ dataset, and at a resolution of approximately 3.25 m for the full dataset. 

The reflectances of the Urban image are then adapted to the topography by using the physical model by Hapke \cite{hapke_theory_2012}, which describes the reflectance of a material as a function of the geometric parameters of the scene and the \textit{single scattering albedo} (SSA) of a material. The SSA only depends on the material and is unaffected by the geometry of the scene. We use a simplified version of the inverted Hapke model (see App. \ref{app: hapkes model data} for a derivation) to calculate the SSA of the five reference EMs. Then, using the forward Hapke model and the pixel-specific geometric parameters, we calculate the EMs in each pixel, which results in pixel-wise EM matrices $\E_n, ~n = 1,2, \ldots, N$. Finally, to generate the image, we multiplied the pixel-wise EM matrices $\E_n$ by the Urban ground truth abundances $\A_\mathrm{gt}$. We added noise with an SNR of 40 dB:
\begin{equation}
    \x_n = \E_n \mathbf{a}_{\mathrm{gt}, n} + \bm{\epsilon}_n,  \quad n=1,2,\ldots, N.
\end{equation}
We do this for both the $100\times100$ sub-image and the full image, to obtain the two datasets for the two experiments.

\subsubsection{Results}

\begin{table*}[t]
\centering
\caption{Unmixing results for the Urban datasets with added topography variability. The best results are highlighted in bold, and the second-best are underlined.}
\label{tab: urban results}
\resizebox{\linewidth}{!}{%
\begin{tabular}{|r|r|ccccccccc|}
\hline
\multirow{2}{*}{\textbf{Size}} & \multirow{2}{*}{\textbf{Metric}} & \multicolumn{9}{c|}{\textbf{Method}} \\
\cline{3-11}
& & \textbf{LMM} & \textbf{SLMM} & \textbf{ELMM(WS)} & \textbf{GLMM} & \textbf{RevNet} & \textbf{ReDSUNN} & \textbf{PGMSU} & \textbf{IP-2LMM} & \textbf{2LMM} \\
\hline \hline
\multirow{3}{*}{$100\times100$} & \textbf{RMSE$_\mathbf{X}$} & 0.7635 & \underline{0.0009} & \textbf{0.0001} & 0.0016 & \texttt{err}$^*$ & 0.0974 & 0.0065 & \underline{0.0009} & \underline{0.0009} \\
& \textbf{RMSE$_\mathbf{A}$} & 0.2870 & 0.0891 & 0.1063 & 0.2463 & \texttt{err}$^*$ & 0.4999 & 0.3407 & \underline{0.0885} & \textbf{0.0719} \\
& \textbf{Time (s)} & 2.86 & \textbf{0.002} & 54.6 & 31.3 & \texttt{err}$^*$ & 62.1 & 73.6 & 51.6 & \underline{0.17} \\
\hline
\multirow{3}{*}{$307\times307$} & \textbf{RMSE$_\mathbf{X}$} & 0.7836 & \underline{0.0015} & \textbf{0.0002} & 0.0018 & \texttt{err}$^*$ & \texttt{err}$^*$ & 0.0048 & \texttt{err}$^*$ & \underline{0.0015} \\
& \textbf{RMSE$_\mathbf{A}$} & 0.3531 & \underline{0.1036} & \underline{0.1036} & 0.2444 & \texttt{err}$^*$ & \texttt{err}$^*$ & 0.3440 & \texttt{err}$^*$ & \textbf{0.0775} \\
& \textbf{Time (s)} & 23.8 & \textbf{0.005} & 309 & 354 & \texttt{err}$^*$ & \texttt{err}$^*$ & 480 & \texttt{err}$^*$ & \underline{3.51} \\
\hline
\end{tabular}
} 
\\[0.5em]
\begin{minipage}{\linewidth}
\footnotesize
$^*$\texttt{err} indicates that the method failed due to a memory-related error
\end{minipage}
\end{table*}

\begin{figure}[t]
    \centering
    \includegraphics[width=0.8\linewidth]{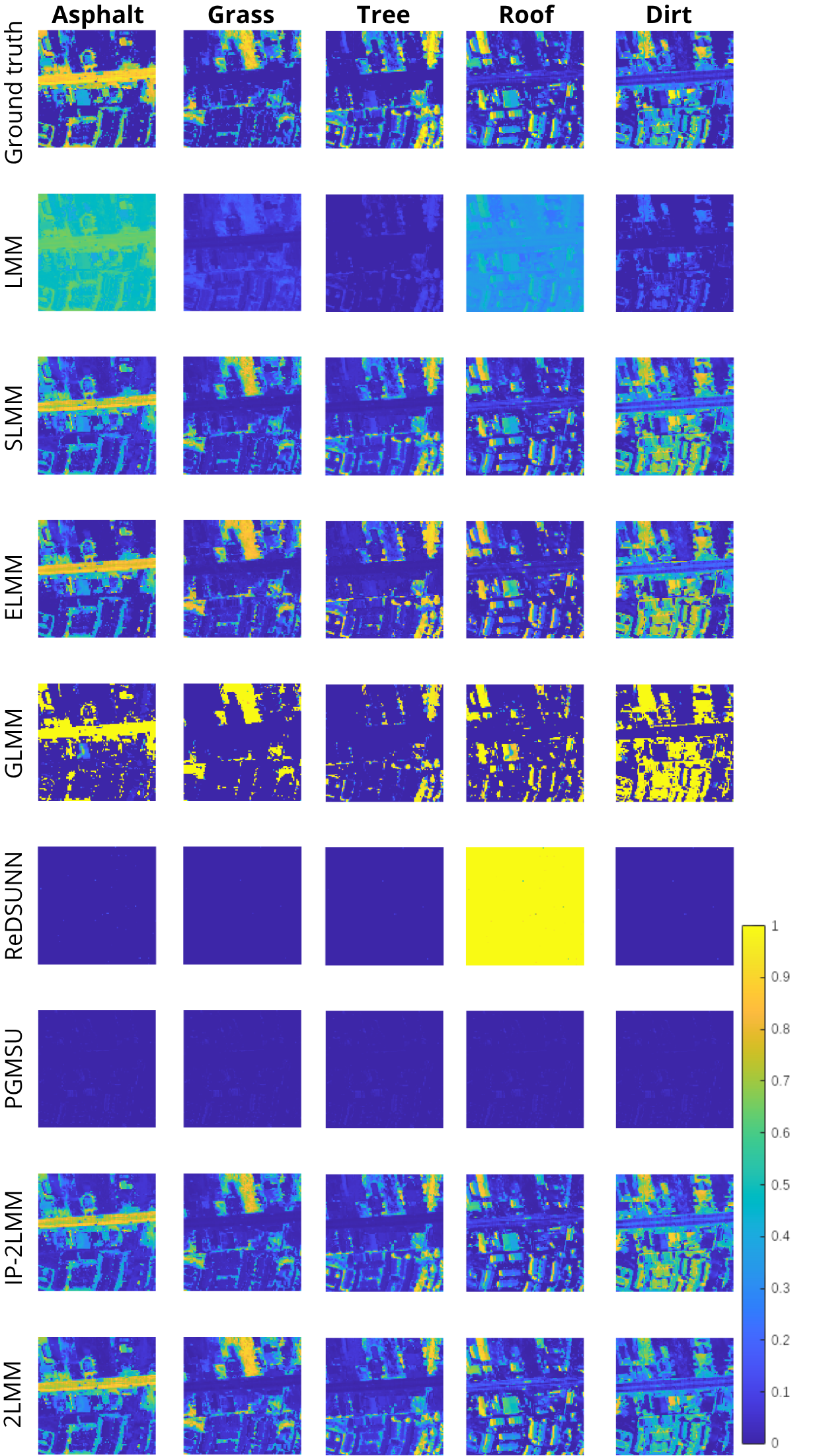}
    \caption{Abundance maps for several methods on the $100\times100$ subimage of the modified Urban dataset. Methods that produced memory errors are omitted.}
    \label{fig: urban abundances}
\end{figure}


\begin{figure}
    \centering
    \includegraphics[width=\linewidth]{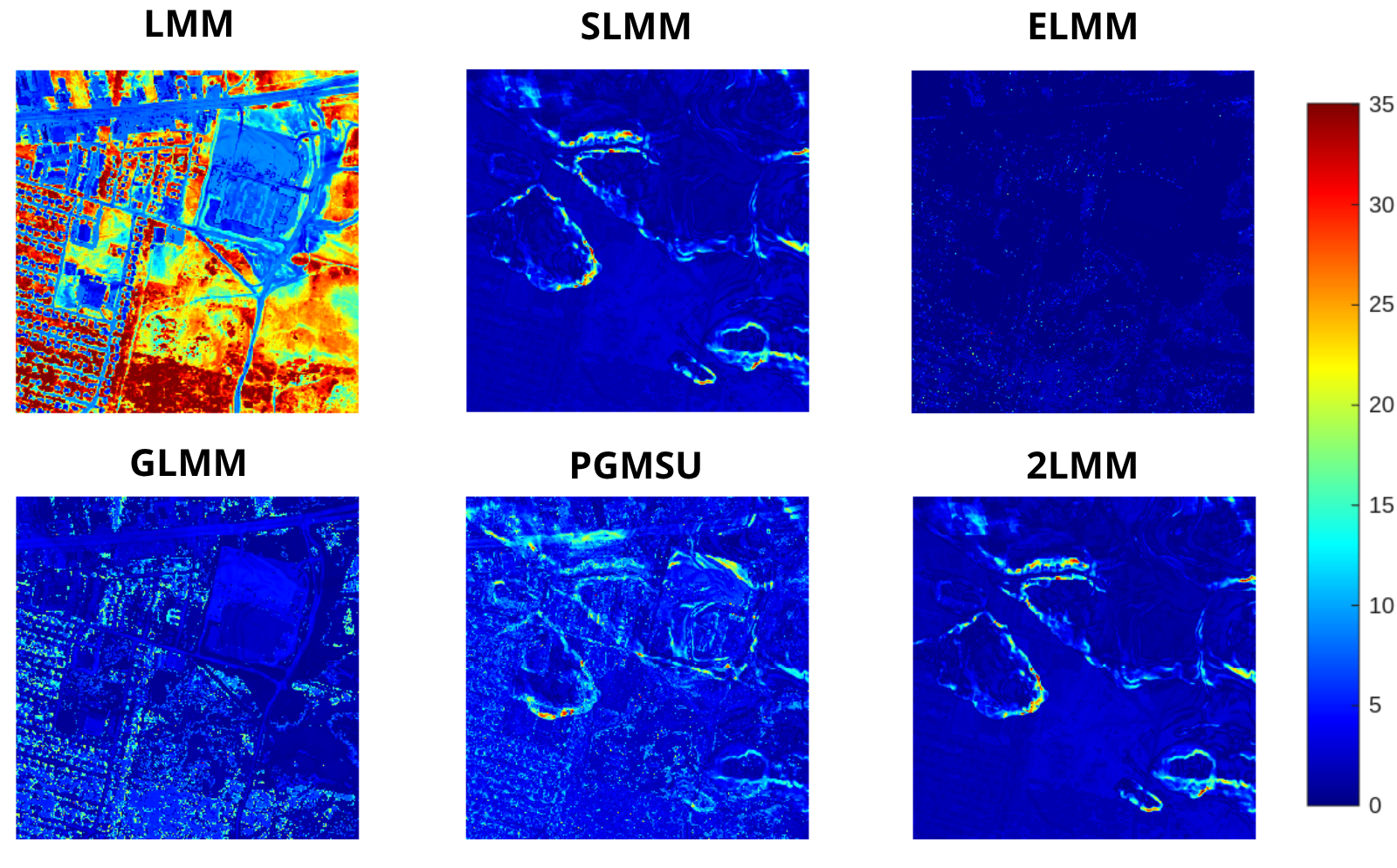}
    \caption{Reconstruction error (RMSE, in \%) for several methods on the full modified Urban dataset. The topography shape is clearly visible in the SLMM and 2LMM error images. Methods that produced memory errors are omitted.}
    \label{fig: urban xl errors}
\end{figure}

During the experimentation, the EMs used for unmixing were obtained using Sisal with a perspective projection applied to the variability-infused images. The topography alters the EM in possibly nonlinear ways, so the EMs provided with the dataset might no longer be valid. We tested the same methods as in the previous experiment for both image sizes. The results are shown in Table \ref{tab: urban results}. For the small image, 2LMM is the best method, in addition to having the second-lowest runtime. The LMM is unable to handle the spectral variability and produces a large error. The abundance maps for the $100 \times 100$ image are shown in Fig. \ref{fig: urban abundances}, and demonstrate that the DL methods fail to produce meaningful abundance maps. We also observed that the final EM estimates of PGMSU and ReDSUNN all collapse to just one EM and discard all others. Furthermore, the GLMM only produces abundance estimates that are close to either zero or one. Since the errors of SLMM and the 2LMM methods are only differing by a small amount, their abundance maps are similar, and are also visually similar to the ground truth. The ELMM abundance maps are also similar to the ground truth, but significant deviations can be observed in highly mixed areas (i.e., areas where no one EM is dominating). This demonstrates that 2LMM performs well in combination with volume-based EEAs, and in scenarios where the model assumptions are not entirely met.

For the full $307\times307$ image, 2LMM remains the best method, again requiring very little time. IP-2LMM fails due to an out-of-memory error, which illustrates its scalability issues. Similarly, two DL methods fail for the full image, and the third (PGMSU) again produces poor results and nonsensical abundance maps. 
We show the pixel-wise reconstruction error for the full dataset in Fig. \ref{fig: urban xl errors}. ELMM shows a near-perfect reconstruction, but we have already established that this does not imply a good abundance estimation. The error maps for SLMM and 2LMM show the contours of the DSM used for generating variability. They originate mostly from steep sections in the DSM, which cause nonlinearities and wavelength-dependent effects. This highlights the limitation of the current version of 2LMM to model these effects.

\section{Conclusion}

In this work, we proposed the 2LMM, a novel physically motivated linear mixing model that mitigates the effect of spectral variability. The model bridges the gap between model complexity and mathematical tractability, which allows for the use of more accurate second-order optimization strategies, boosting the performance across synthetic and real-world datasets, including a new benchmark dataset. We proposed a more scalable version of 2LMM unmixing, compared to the IP method that we proposed previously. This method is based on the limited memory BFGS algorithm, which is a quasi-second-order method. This method achieves performance similar to the full second-order IP method while drastically reducing runtime and memory consumption. This makes the 2LMM a highly efficient and practical choice for unmixing under variability that can outperform the state of the art, including model-based and DL methods.

The new 2LMM method still struggles with modeling nonlinear effects and still requires more memory than a standard Alternating Least Squares approach. Secondly, the method does not adaptively adjust the EMs during optimization. Therefore, future work will focus on designing more memory-efficient algorithms for 2LMM unmixing. Furthermore, extensions to joint EM and abundance estimation and to nonlinear scenarios will be investigated.

\bibliographystyle{IEEEtran}
\bibliography{main}

@article{lee_learning_1999,
	title = {Learning the parts of objects by non-negative matrix factorization},
	volume = {401},
	copyright = {1999 Macmillan Magazines Ltd.},
	issn = {1476-4687},
	 
	doi = {10.1038/44565},
	language = {en},
	number = {6755},
	urldate = {2024-09-12},
	journal = {Nature},
	author = {Lee, Daniel D. and Seung, H. Sebastian},
	month = oct,
	year = {1999},
	pages = {788--791},
}

@article{drumetz_spectral_2020,
	title = {Spectral {Variability} {Aware} {Blind} {Hyperspectral} {Image} {Unmixing} {Based} on {Convex} {Geometry}},
	volume = {29},
	issn = {1941-0042},
	 
	doi = {10.1109/TIP.2020.2974062},
	urldate = {2024-09-06},
	journal = {IEEE Trans. Image Process.},
	author = {Drumetz, Lucas and Chanussot, Jocelyn and Jutten, Christian and Ma, Wing-Kin and Iwasaki, Akira},
	year = {2020},
	pages = {4568--4582},
}

@article{drumetz_blind_2016,
	title = {Blind {Hyperspectral} {Unmixing} {Using} an {Extended} {Linear} {Mixing} {Model} to {Address} {Spectral} {Variability}},
	volume = {25},
	issn = {1941-0042},
	 
	doi = {10.1109/TIP.2016.2579259},
	number = {8},
	urldate = {2024-09-06},
	journal = {IEEE Trans. Image Process.},
	author = {Drumetz, Lucas and Veganzones, Miguel-Angel and Henrot, Simon and Phlypo, Ronald and Chanussot, Jocelyn and Jutten, Christian},
	month = aug,
	year = {2016},
	pages = {3890--3905},
}

@article{yuan_projection-based_2015,
	title = {Projection-{Based} {NMF} for {Hyperspectral} {Unmixing}},
	volume = {8},
	issn = {2151-1535},
	 
	doi = {10.1109/JSTARS.2015.2427656},
	number = {6},
	urldate = {2024-09-06},
	journal = {IEEE J. Sel. Top. Appl. Earth Obs. Remote Sens.},
	author = {Yuan, Yuan and Feng, Yachuang and Lu, Xiaoqiang},
	month = jun,
	year = {2015},
	pages = {2632--2643},
}

@article{shaw_spectral_2003,
	title = {Spectral {Imaging} for {Remote} {Sensing}},
	volume = {14},
	journal = {Lincoln Laboratory Journal},
	author = {Shaw, Gary and Burke, Hsiao-hua},
	month = jan,
	year = {2003},
}

@article{heylen_review_2014,
	title = {A {Review} of {Nonlinear} {Hyperspectral} {Unmixing} {Methods}},
	volume = {7},
	issn = {2151-1535},
	 
	doi = {10.1109/JSTARS.2014.2320576},
	number = {6},
	urldate = {2024-09-05},
	journal = {IEEE J. Sel. Top. Appl. Earth Obs. Remote Sens.},
	author = {Heylen, Rob and Parente, Mario and Gader, Paul},
	month = jun,
	year = {2014},
	pages = {1844--1868},
}

@article{borsoi_spectral_2021,
	title = {Spectral {Variability} in {Hyperspectral} {Data} {Unmixing}: {A} comprehensive review},
	volume = {9},
	issn = {2168-6831},
	shorttitle = {Spectral {Variability} in {Hyperspectral} {Data} {Unmixing}},
	 
	doi = {10.1109/MGRS.2021.3071158},
	number = {4},
	urldate = {2024-09-04},
	journal = {IEEE Geosci. Remote Sens. Mag.},
	author = {Borsoi, Ricardo Augusto and Imbiriba, Tales and Bermudez, José Carlos Moreira and Richard, Cédric and Chanussot, Jocelyn and Drumetz, Lucas and Tourneret, Jean-Yves and Zare, Alina and Jutten, Christian},
	month = dec,
	year = {2021},
	pages = {223--270},
}

@article{hong_augmented_2019,
	title = {An {Augmented} {Linear} {Mixing} {Model} to {Address} {Spectral} {Variability} for {Hyperspectral} {Unmixing}},
	volume = {28},
	issn = {1941-0042},
	 
	doi = {10.1109/TIP.2018.2878958},
	number = {4},
	urldate = {2024-10-11},
	journal = {IEEE Trans. Image Process.},
	author = {Hong, Danfeng and Yokoya, Naoto and Chanussot, Jocelyn and Zhu, Xiao Xiang},
	month = apr,
	year = {2019},
	pages = {1923--1938},
}

@inproceedings{bioucas-dias_alternating_2010,
	title = {Alternating direction algorithms for constrained sparse regression: {Application} to hyperspectral unmixing},
	shorttitle = {Alternating direction algorithms for constrained sparse regression},
	 
	doi = {10.1109/WHISPERS.2010.5594963},
	urldate = {2024-10-30},
	booktitle = {2010 2nd {Workshop} on {Hyperspectral} {Image} and {Signal} {Processing}: {Evolution} in {Remote} {Sensing}},
	author = {Bioucas-Dias, José M. and Figueiredo, Mario A. T.},
	month = jun,
	year = {2010},
	pages = {1--4},
}

@book{nocedal_numerical_2006,
	series = {{Operations} {Research} and {Financial} {Engineering}},
	title = {Numerical {Optimization}},
	copyright = {http://www.springer.com/tdm},
	isbn = {978-0-387-30303-1},
	 
	language = {en},
	urldate = {2024-10-30},
	publisher = {Springer (New York)},
	author = {Nocedal, Jorge and Wright, Stephen J.},
	year = {2006},
}

@inproceedings{imbiriba_generalized_2018,
	title = {Generalized {Linear} {Mixing} {Model} {Accounting} for {Endmember} {Variability}},
	 
	doi = {10.1109/ICASSP.2018.8462214},
	urldate = {2024-10-30},
	booktitle = {2018 {IEEE} {International} {Conference} on {Acoustics}, {Speech} and {Signal} {Processing} ({ICASSP})},
	author = {Imbiriba, Tales and Borsoi, Ricardo Augusto and Moreira Bermudez, José Carlos},
	month = apr,
	year = {2018},
	pages = {1862--1866},
}

@book{hapke_theory_2012,
	address = {Cambridge},
	edition = {2nd},
	title = {Theory of {Reflectance} and {Emittance} {Spectroscopy}},
	isbn = {978-0-521-88349-8},
	urldate = {2024-10-31},
	publisher = {Cambridge University Press},
	author = {Hapke, Bruce},
	year = {2012},
	doi = {10.1017/CBO9781139025683},
}

@phdthesis{kozintsev_computations_1999,
	address = {College Park},
	type = {{PhD} {Thesis}},
	title = {Computations with {Gaussian} {Random} {Fields}},
	school = {University of Maryland},
	author = {Kozintsev, Boris},
	year = {1999},
}

@techreport{kokaly_usgs_2017,
	title = {{USGS} {Spectral} {Library} {Version} 7},
	language = {en},
	number = {1035},
	urldate = {2024-11-20},
	institution = {U.S. Geological Survey},
	author = {Kokaly, Raymond F. and Clark, Roger N. and Swayze, Gregg A. and others},
    year = {2017},
}

@article{keshava_spectral_2002,
	title = {Spectral unmixing},
	volume = {19},
	issn = {1558-0792},
	 
	doi = {10.1109/79.974727},
	number = {1},
	urldate = {2024-12-05},
	journal = {IEEE Signal Process. Mag.},
	author = {Keshava, N. and Mustard, J.F.},
	month = jan,
	year = {2002},
	pages = {44--57},
}

@article{nascimento_vertex_2005,
	title = {Vertex component analysis: a fast algorithm to unmix hyperspectral data},
	volume = {43},
	issn = {1558-0644},
	shorttitle = {Vertex component analysis},
	 
	doi = {10.1109/TGRS.2005.844293},
	number = {4},
	urldate = {2024-12-19},
	journal = {IEEE Trans. Geosci. Remote Sens. },
	author = {Nascimento, J.M.P. and Dias, J.M.B.},
	month = apr,
	year = {2005},
	pages = {898--910},
}

@inproceedings{haijen_two-step_2025,
	title = {A {Two}-{Step} {Linear} {Mixing} {Model} for {Unmixing} {Under} {Hyperspectral} {Variability}},
	doi = {10.1109/IGARSS55030.2025.11243310},
	urldate = {2025-12-01},
	booktitle = {2025 {IEEE} {Int.} {Geosci.} {Remote} {Sens.} {Symp.}},
	author = {Haijen, Xander and Koirala, Bikram and Tao, Xuanwen and Scheunders, Paul},
	month = aug,
	year = {2025},
	pages = {8707--8711},
}

@article{roberts_mapping_1998,
	title = {Mapping {Chaparral} in the {Santa} {Monica} {Mountains} {Using} {Multiple} {Endmember} {Spectral} {Mixture} {Models}},
	volume = {65},
	issn = {0034-4257},
	doi = {10.1016/S0034-4257(98)00037-6},
	number = {3},
	urldate = {2025-01-14},
	journal = {Remote Sens. Environ.},
	author = {Roberts, D. A. and Gardner, M. and Church, R. and Ustin, S. and Scheer, G. and Green, R. O.},
	month = sep,
	year = {1998},
	pages = {267--279},
}

@article{iordache_total_2012,
	title = {Total {Variation} {Spatial} {Regularization} for {Sparse} {Hyperspectral} {Unmixing}},
	volume = {50},
	issn = {1558-0644},
	 
	doi = {10.1109/TGRS.2012.2191590},
	number = {11},
	urldate = {2025-01-14},
	journal = {IEEE Trans. Geosci. Remote Sens. },
	author = {Iordache, Marian-Daniel and Bioucas-Dias, José M. and Plaza, Antonio},
	month = nov,
	year = {2012},
	pages = {4484--4502},
}

@article{song_weighted_2022,
	title = {Weighted {Total} {Variation} {Regularized} {Blind} {Unmixing} for {Hyperspectral} {Image}},
	volume = {19},
	issn = {1558-0571},
	 
	doi = {10.1109/LGRS.2021.3094826},
	urldate = {2025-01-14},
	journal = {IEEE Geoscience and Remote Sensing Letters},
	author = {Song, Hanjie and Wu, Xing and Zou, Anqi and Liu, Yang and Zou, Yongliao},
	year = {2022},
	pages = {1--5},
}

@article{chen_dsfc-ae_2024,
	title = {{DSFC}-{AE}: {A} {New} {Hyperspectral} {Unmixing} {Method} {Based} on {Deep} {Shared} {Fully} {Connected} {Autoencoder}},
	volume = {17},
	issn = {2151-1535},
	shorttitle = {{DSFC}-{AE}},
	 
	doi = {10.1109/JSTARS.2024.3450856},
	urldate = {2025-01-24},
	journal = {IEEE J. Sel. Top. Appl. Earth Obs. Remote Sens.},
	author = {Chen, Hao and Chen, Tao and Zhang, Yuxiang and Du, Bo and Plaza, Antonio},
	year = {2024},
	pages = {15746--15760},
}

@article{zheng_blind_2024,
	title = {Blind {Unmixing} {Using} {Dispersion} {Model}-{Based} {Autoencoder} to {Address} {Spectral} {Variability}},
	volume = {62},
	issn = {1558-0644},
	 
	doi = {10.1109/TGRS.2024.3399003},
	urldate = {2025-01-24},
	journal = {IEEE Trans. Geosci. Remote Sens. },
	author = {Zheng, Haoren and Li, Zulong and Sun, Chenyu and Zhang, Hanqiu and Liu, Hongyi and Wei, Zhihui},
	year = {2024},
	pages = {1--14},
}

@article{shi_deep_2022,
	title = {Deep {Generative} {Model} for {Spatial}–{Spectral} {Unmixing} {With} {Multiple} {Endmember} {Priors}},
	volume = {60},
	issn = {1558-0644},
	 
	doi = {10.1109/TGRS.2022.3168712},
	urldate = {2025-01-24},
	journal = {IEEE Trans. Geosci. Remote Sens. },
	author = {Shi, Shuaikai and Zhang, Lijun and Altmann, Yoann and Chen, Jie},
	year = {2022},
	pages = {1--14},
}

@article{zhang_spectral_2022,
	title = {Spectral {Variability} {Augmented} {Two}-{Stream} {Network} for {Hyperspectral} {Sparse} {Unmixing}},
	volume = {19},
	issn = {1558-0571},
	 
	doi = {10.1109/LGRS.2022.3214843},
	urldate = {2025-01-24},
	journal = {IEEE Geoscience and Remote Sensing Letters},
	author = {Zhang, Ge and Mei, Shaohui and Xie, Bobo and Feng, Yan and Du, Qian},
	year = {2022},
	pages = {1--5},
}

@article{su_multi-attention_2023,
	title = {A {Multi}-{Attention} {Autoencoder} for {Hyperspectral} {Unmixing} {Based} on the {Extended} {Linear} {Mixing} {Model}},
	volume = {15},
	copyright = {http://creativecommons.org/licenses/by/3.0/},
	issn = {2072-4292},
	 
	doi = {10.3390/rs15112898},
	language = {en},
	number = {11},
	urldate = {2025-01-24},
	journal = {Remote Sens.},
	author = {Su, Lijuan and Liu, Jun and Yuan, Yan and Chen, Qiyue},
	month = jan,
	year = {2023},
	pages = {2898},
}

@article{cheng_hyperspectral_2023,
	title = {Hyperspectral {Unmixing} {Network} {Accounting} for {Spectral} {Variability} {Based} on a {Modified} {Scaled} and a {Perturbed} {Linear} {Mixing} {Model}},
	volume = {15},
	copyright = {http://creativecommons.org/licenses/by/3.0/},
	issn = {2072-4292},
	 
	doi = {10.3390/rs15153890},
	language = {en},
	number = {15},
	urldate = {2025-01-24},
	journal = {Remote Sens.},
	author = {Cheng, Ying and Zhao, Liaoying and Chen, Shuhan and Li, Xiaorun},
	month = jan,
	year = {2023},
	pages = {3890},
}

@article{hong_endmember-guided_2022,
	title = {Endmember-{Guided} {Unmixing} {Network} ({EGU}-{Net}): {A} {General} {Deep} {Learning} {Framework} for {Self}-{Supervised} {Hyperspectral} {Unmixing}},
	volume = {33},
	issn = {2162-2388},
	shorttitle = {Endmember-{Guided} {Unmixing} {Network} ({EGU}-{Net})},
	 
	doi = {10.1109/TNNLS.2021.3082289},
	number = {11},
	urldate = {2025-01-24},
	journal = {IEEE Trans. Neural Netw. Learn. Syst.},
	author = {Hong, Danfeng and Gao, Lianru and Yao, Jing and Yokoya, Naoto and Chanussot, Jocelyn and Heiden, Uta and Zhang, Bing},
	month = nov,
	year = {2022},
	pages = {6518--6531},
}

@article{gao_reversible_2024,
	title = {A {Reversible} {Generative} {Network} for {Hyperspectral} {Unmixing} {With} {Spectral} {Variability}},
	volume = {62},
	issn = {1558-0644},
	 
	doi = {10.1109/TGRS.2024.3403926},
	urldate = {2025-01-24},
	journal = {IEEE Trans. Geosci. Remote Sens. },
	author = {Gao, Yuyou and Pan, Bin and Xu, Xia and Song, Xinyu and Shi, Zhenwei},
	year = {2024},
	pages = {1--15},
}

@article{gao_proportional_2024,
	title = {Proportional {Perturbation} {Model} for {Hyperspectral} {Unmixing} {Accounting} for {Endmember} {Variability}},
	volume = {21},
	issn = {1558-0571},
	 
	doi = {10.1109/LGRS.2024.3350889},
	urldate = {2025-01-24},
	journal = {IEEE Geosci. Remote Sens. Let.},
	author = {Gao, Wei and Yang, Jingyu and Chen, Jie},
	year = {2024},
	pages = {1--5},
}

@article{borsoi_dynamical_2023,
	title = {Dynamical {Hyperspectral} {Unmixing} {With} {Variational} {Recurrent} {Neural} {Networks}},
	volume = {32},
	issn = {1941-0042},
	 
	doi = {10.1109/TIP.2023.3266660},
	urldate = {2025-01-24},
	journal = {IEEE Trans. Image Process.},
	author = {Borsoi, Ricardo A. and Imbiriba, Tales and Closas, Pau},
	year = {2023},
	pages = {2279--2294},
}

@article{sun_blind_2022,
	title = {Blind {Unmixing} of {Hyperspectral} {Images} {Based} on {L1} {Norm} and {Tucker} {Tensor} {Decomposition}},
	volume = {19},
	issn = {1558-0571},
	 
	doi = {10.1109/LGRS.2021.3103962},
	urldate = {2025-01-30},
	journal = {IEEE Geoscience and Remote Sensing Letters},
	author = {Sun, Le and Guo, Huxiang},
	year = {2022},
	pages = {1--5},
}

@inproceedings{winter_n-findr_1999,
	title = {N-{FINDR}: an algorithm for fast autonomous spectral end-member determination in hyperspectral data},
	volume = {3753},
	shorttitle = {N-{FINDR}},
	 
	doi = {10.1117/12.366289},
	urldate = {2025-02-11},
	booktitle = {Imaging {Spectrometry} {V}},
	publisher = {SPIE},
	author = {Winter, Michael E.},
	month = oct,
	year = {1999},
	pages = {266--275},
}

@article{shi_probabilistic_2022,
	title = {Probabilistic {Generative} {Model} for {Hyperspectral} {Unmixing} {Accounting} for {Endmember} {Variability}},
	volume = {60},
	issn = {1558-0644},
	 
	doi = {10.1109/TGRS.2021.3121799},
	urldate = {2025-02-13},
	journal = {IEEE Trans. Geosci. Remote Sens. },
	author = {Shi, Shuaikai and Zhao, Min and Zhang, Lijun and Altmann, Yoann and Chen, Jie},
	year = {2022},
	pages = {1--15},
}

@article{uezato_hierarchical_2020,
	title = {Hierarchical {Sparse} {Nonnegative} {Matrix} {Factorization} for {Hyperspectral} {Unmixing} with {Spectral} {Variability}},
	volume = {12},
	copyright = {http://creativecommons.org/licenses/by/3.0/},
	issn = {2072-4292},
	 
	doi = {10.3390/rs12142326},
	language = {en},
	number = {14},
	urldate = {2025-02-28},
	journal = {Remote Sensing},
	author = {Uezato, Tatsumi and Fauvel, Mathieu and Dobigeon, Nicolas},
	month = jan,
	year = {2020},
	pages = {2326},
}

@article{thouvenin_hyperspectral_2016,
	title = {Hyperspectral {Unmixing} {With} {Spectral} {Variability} {Using} a {Perturbed} {Linear} {Mixing} {Model}},
	volume = {64},
	issn = {1941-0476},
	 
	doi = {10.1109/TSP.2015.2486746},
	number = {2},
	urldate = {2025-03-04},
	journal = {IEEE Trans. Signal Process.},
	author = {Thouvenin, Pierre-Antoine and Dobigeon, Nicolas and Tourneret, Jean-Yves},
	month = jan,
	year = {2016},
	pages = {525--538},
}

@inproceedings{bioucas-dias_variable_2009,
	title = {A variable splitting augmented {Lagrangian} approach to linear spectral unmixing},
	 
	doi = {10.1109/WHISPERS.2009.5289072},
	urldate = {2025-03-04},
	booktitle = {1st Worksh. Hypersp. Image Signal Process. Evol. Remote Sens.},
	author = {Bioucas-Dias, Jose M.},
	month = aug,
	year = {2009},
	pages = {1--4},
}

@article{zhuang_regularization_2019,
	title = {Regularization {Parameter} {Selection} in {Minimum} {Volume} {Hyperspectral} {Unmixing}},
	volume = {57},
	issn = {1558-0644},
	 
	doi = {10.1109/TGRS.2019.2929776},
	number = {12},
	urldate = {2025-03-06},
	journal = {IEEE Trans. Geosci. Remote Sens. },
	author = {Zhuang, Lina and Lin, Chia-Hsiang and Figueiredo, Mário A. T. and Bioucas-Dias, José M.},
	month = dec,
	year = {2019},
	pages = {9858--9877},
}

@article{fu_identifiability_2018,
	title = {On {Identifiability} of {Nonnegative} {Matrix} {Factorization}},
	volume = {25},
	issn = {1558-2361},
	 
	doi = {10.1109/LSP.2018.2789405},
	number = {3},
	urldate = {2025-03-06},
	journal = {IEEE Signal Process. Let.},
	author = {Fu, Xiao and Huang, Kejun and Sidiropoulos, Nicholas D.},
	month = mar,
	year = {2018},
	pages = {328--332},
}

@article{fu_nonnegative_2019,
	title = {Nonnegative {Matrix} {Factorization} for {Signal} and {Data} {Analytics}: {Identifiability}, {Algorithms}, and {Applications}},
	volume = {36},
	issn = {1558-0792},
	shorttitle = {Nonnegative {Matrix} {Factorization} for {Signal} and {Data} {Analytics}},
	 
	doi = {10.1109/MSP.2018.2877582},
	number = {2},
	urldate = {2025-03-10},
	journal = {IEEE Signal Process. Mag.},
	author = {Fu, Xiao and Huang, Kejun and Sidiropoulos, Nicholas D. and Ma, Wing-Kin},
	month = mar,
	year = {2019},
	pages = {59--80},
}

@article{fu_robust_2016,
	title = {Robust {Volume} {Minimization}-{Based} {Matrix} {Factorization} for {Remote} {Sensing} and {Document} {Clustering}},
	volume = {64},
	issn = {1941-0476},
	   doi = {10.1109/TSP.2016.2602800},
	number = {23},
	urldate = {2025-03-10},
	journal = {IEEE Trans. Signal Process.},
	author = {Fu, Xiao and Huang, Kejun and Yang, Bo and Ma, Wing-Kin and Sidiropoulos, Nicholas D.},
	month = dec,
	year = {2016},
	pages = {6254--6268},
}

@inproceedings{boardman_mapping_1995,
	title = {Mapping target signatures via partial unmixing of {AVIRIS} data},
	urldate = {2025-03-24},
	author = {Boardman, Joseph W. and Kruse, Fred A. and Green, Robert O.},
    booktitle = {Fifth Annual JPL Airborne Earth Science Workshop},
	month = jan,
	year = {1995},
}

@article{de_sterck_nonlinearly_2018,
	title = {Nonlinearly preconditioned {L}-{BFGS} as an acceleration mechanism for alternating least squares with application to tensor decomposition},
	volume = {25},
	issn = {1099-1506},
	 
	doi = {10.1002/nla.2202},
	language = {en},
	number = {6},
	urldate = {2025-03-31},
	journal = {Numer. Linear Algebra Appl.},
	author = {De Sterck, Hans and Howse, Alexander J.M.},
	year = {2018},
}

@inproceedings{haijen_benchmark_2025,
	title = {A {Benchmark} {Linear} {Unmixing} {Dataset} with {Spectral} {Variability} and {Ground} {Truth}},
	author = {Haijen, Xander and Koirala, Bikram and Tao, Xuanwen and Scheunders, Paul},
	month = nov,
	year = {2025},
	note = {{T}o appear at 2025 {Workshop} {Hypersp.} {Image} {Signal} {Process.}: {Evol.} {Remote} {Sens.} (WHISPERS)}
}

@article{liu_limited_1989,
	title = {On the limited memory {BFGS} method for large scale optimization},
	volume = {45},
	issn = {1436-4646},
	 
	doi = {10.1007/BF01589116},
	language = {en},
	number = {1},
	urldate = {2025-10-16},
	journal = {Math. Prog.},
	author = {Liu, Dong C. and Nocedal, Jorge},
	month = aug,
	year = {1989},
	pages = {503--528},
}

{\appendices

\section{Data generation using the Hapke model} \label{app: hapkes model data}
We ignore anisotropic scattering and the opposition effects to simplify the Hapke model (see \cite{heylen_review_2014} for a concise description of these effects). This leads to the wavelength-dependent model:
\begin{equation}
    x(\lambda) = \frac{w(\lambda) \mu_0}{4\pi (\mu_0 + \mu)} H(w(\lambda), \mu) H(w(\lambda), \mu_0)
\end{equation}
where $\mu = \cos \theta_r$ and $\mu_0 = \cos \theta_0$ are the cosines of the reflected and incident angles, respectively, measured between the surface normal and the line from the sensor (resp. source) to the sample. $w(\lambda)$ is the single-scattering albedo (SSA) at wavelength $\lambda$. Lastly, $H(w(\lambda), \mu)$ is Chandrasekhar’s isotropic scattering function, which can be approximated by \cite{heylen_review_2014}:
\begin{equation}
    H(w(\lambda), \mu) = \frac{1+2\mu}{1+2\mu \sqrt{1-w(\lambda)}}.
\end{equation}
We can convert this to a reflectance relative to a reference white panel ($w = 1$):
\begin{equation} \label{eq: relative hapke model}
    \begin{aligned}
            y(\lambda) &= \frac{w(\lambda) H(w(\lambda), \mu) H(w(\lambda), \mu_0)}{H(1, \mu)H(1, \mu_0)} \\
            &= \frac{w(\lambda)}{\left(1+2\mu\sqrt{1-w(\lambda)}\right)\left(1+2\mu_0\sqrt{1-w(\lambda)}\right)}.        
    \end{aligned}
\end{equation}
For simplicity of notation, the equation is written for wavelength-dependent scalars. It can be applied element-wise to the entire spectrum. To obtain the SSAs of the materials, the formula (\ref{eq: relative hapke model}) can be inverted and rewritten for $w(\lambda)$. This way, we can obtain the SSAs from measured reflectances, and then use the SSAs to generate the pixel-wise EMs by using (\ref{eq: relative hapke model}) and filling in the values for $\mu$ and $\mu_0$ in every pixel, which are derived from the DSM.
}

\newpage

\vfill

\end{document}